\documentclass[12pt]{article}
\usepackage{amssymb,amsmath,epsfig}
\allowdisplaybreaks

\begin{document}

\title{\bf Viable Wormhole Solutions in Energy-Momentum Squared Gravity}
\author{M. Sharif \thanks {msharif.math@pu.edu.pk} and
M. Zeeshan Gul \thanks{mzeeshangul.math@gmail.com}\\
Department of Mathematics, University of the Punjab,\\
Quaid-e-Azam Campus, Lahore-54590, Pakistan.}

\date{}
\maketitle

\begin{abstract}
This paper investigates static wormhole solutions through Noether
symmetry approach in the context of energy-momentum squared gravity.
This newly developed proposal resolves the singularity of big-bang
and yields feasible cosmological results in the early times. We
consider the particular model of this theory to establish symmetry
generators and corresponding conserved quantities. For constant and
variable red-shift functions, we examine the presence of viable
traversable wormhole solutions for both dust as well as non-dust
matter distributions and analyze the stable state of these
solutions. We investigate the graphical interpretation of null and
weak energy bounds for normal and effective energy-momentum tensors
to examine the presence of physically viable wormhole geometry. It
is found that realistic traversable and stable wormhole solutions
are obtained for a particular model of this gravity.
\end{abstract}
Wormhole solutions.

\section{Introduction}

The accelerated expansion of the universe has been the most stunning
and dazzling consequence for scientific community over the past two
decades. This expansion is considered as the result of some
ambiguous force dubbed as dark energy which has repulsive effects.
This cryptic energy has motivated many researchers to reveal its
hidden characteristics which are still unknown. In this perspective,
modified theories of gravity are known as the most significant and
elegant proposals to unveil the cosmic mysteries. These proposals
can be established by introducing the curvature invariant and their
corresponding functions in the curvature part of the
Einstein-Hilbert action. The $f(R)$ theory is the simplest
modification of general relativity (GR). The significant literature
\cite{1} has been accessible to understand the viable attributes of
this modified theory.

The curvature-matter coupled theories have become the subject of
great interest for cosmologists due to the interactions among the
geometric and matter part. These interactions determine the distinct
stages of the universe and the rotation curves of galaxies. The
conservation law does not hold in these theories that yield the
presence of an additional force. Such theories are very helpful to
understand the cosmic acceleration as well as interactions between
the dark components. Harko et al. \cite{3} developed such
interactions in $f(R)$ gravity named as $f(R,T)$ theory. The
non-minimally interaction of curvature with matter was established
in \cite{4}, named as $f(R, T,R_{\alpha\beta} T^{\alpha\beta})$
theory. One such coupling yields $f(R, T^{\phi})$ theory \cite{5}.

The existence of singularities in GR is a critical issue due to its
prediction at high energy regime, where GR is not applicable because
of the expected quantum effects. Nevertheless, there is no
particular technique for quantum theory. Accordingly,
energy-momentum squared gravity (EMSG) (also known as
$f(R,\mathbf{T}^{2})$ gravity) has been established by incorporating
the analytic function $T_{\alpha\beta}T^{\alpha\beta}$ in the
generic action where $\mathbf{T}^{2}$ is denoted by
$T_{\alpha\beta}T^{\alpha\beta}$. \cite{6}. It provides squared
terms of the fluid variables and their products in the equations of
motion which help to explain different captivating cosmological
results. This theory has a regular bounce with finite maximum energy
density and a minimum scale factor at early times. As a result, it
can resolve big-bang singularity with a non-quantum prescription. It
is mentioned here that this proposal resolves the spacetime
singularity but cosmological evolution remains unaffected.

Further work on this proposal has been carried out by many
researchers \cite{6a}. Board and Barrow \cite{7} analyzed the
analytic solutions for the isotropic universe and examined their
actions with cosmic expansion, existence and avoidance of
singularities. Nari and Roshan \cite{8} investigated the physically
realistic and stable dense objects. Morares and Sahoo \cite{9}
examined non-exotic matter wormholes in this background. Bahamonde
et al. \cite{11} explored minimal as well as non-minimal coupling
models of EMSG and found that these models explain expanding
behavior of the universe. Recently, we have studied the Noether
symmetry approach in this framework and examined the physically
viable solutions through different cosmological parameters. We have
also studied the viability and stability of dense objects. It is
found that modified EMSG terms boost the stability of system and
hence prevent the collapse rate \cite{12}. It is clear from the
aforementioned references that EMSG needs more attention and
therefore motivation to investigate such a theory is very strong.
There are several open problems that may be explored and this will
upgrade our knowledge about various alternative gravitational
theories.

Symmetry is a familiar and important ingredient of cosmology and
theoretical physics. In this perspective, the Noether symmetry
strategy is supposed as the most fascinating method that exhibits a
relation among symmetry generators and conserved quantities of a
dynamical system \cite{14}. These symmetries are very helpful to
establish the exact solution of a nonlinear system by minimizing
them to a linear one. Capozziello et al. \cite{15} found the exact
solutions of static and non-static spherical spacetime via the
Noether symmetry technique in $f(R)$ gravity. Shamir et al.
\cite{16} used this strategy to investigate the stability of
spherically symmetric and Friedmann-Robertson-Walker universe in the
same theory. Kucukakca et al. \cite{17} studied exact solutions of
the Bianchi type-I universe via the Noether symmetry technique.
Sharif and his collaborators \cite{19} examined cosmic expansion and
evolution by using this strategy.

Our universe puts forward stunning questions for the researchers due
to its surprising and enigmatic nature. The presence of hypothetical
structures is viewed as the most controversial problem that yields
the structure of a wormhole (WH). It is defined as a speculative
tunnel that joins two different regions of spacetimes in the
presence of exotic matter. If a hypothetical bridge joins distinct
sectors of the same universe then intra-universe WH appears while
for two different spacetimes inter-universe WH exists. The
appearance of a physically realistic WH is questioned due to large
amount of exotic matter. Hence, for a physically viable WH geometry,
the exotic matter in the bridge must be minimum. Apart from the
presence of such astrophysical geometries, stability analysis is the
most critical issue which determines their actions against
perturbations and boosts the physical characterization. The
configuration without singularity demonstrates a stable state that
restricts the WH to collapse whereas unstable WH may also exist due
to very slow decay. The evolution of system instability may
contribute to several phenomena of interest from structure formation
to supernova explosions. To investigate WH geometry, different
techniques have been proposed to examine the presence of physically
viable WH geometry \cite{20}.

In modified gravitational theories, the study of WH geometry has
been incredibly enthusiastic for cosmologists. Bahamonde et al.
\cite{22} formulated physically realistic WH solutions for
Friedmann-Robertson-Walker spacetime in $f(R)$ theory. Sharif and
Fatima \cite{23} examined the static and non-static WH solutions in
$f(\mathcal{G})$ gravity. Mazharimousavi and Halilsoy \cite{24}
studied the solution of WH structure near the throat that fulfills
all the required WH conditions for both vacuum/non-vacuum cases in
the framework of $f(R)$ theory. Bahamonde et al. \cite{25} applied
the Noether symmetry technique to derive the physically viable and
traversable WH solutions in the background of scalar-tensor theory.
Sharif and Nawazish \cite{26} formulated static WH solutions via the
Noether symmetry technique and found the stable state of WH for both
constant/variable red-shift function in $f(R)$ theory. Zubair et al.
\cite{27} investigated the presence of static WH geometry with
various matter configurations in $f(R,T)$ gravity.

In this paper, we use the Noether symmetry technique to analyze the
geometry of WH for both dust as well as non-dust matter distribution
in EMSG. The paper is organized as follows. In section \textbf{2},
we establish the field equations of static spherical system and
energy bounds in the background of EMSG. Section \textbf{3} is
devoted to formulate point-like Lagrangian. Section \textbf{4}
provides brief information of WH solutions via the Noether symmetry
technique for a particular EMSG model and analyze the physical
presence through energy conditions graphically. In section
\textbf{5}, we investigate the stability of WH solutions by
Tolman-Oppenheimer-Volkov (TOV) equation. A brief description and
discussion of the outcomes are bestowed in the last section.

\section{Energy-Momentum Squared Gravity}

We establish the equations of motion with isotropic matter
distribution in this section. The action for this gravity is
determined as \cite{6}
\begin{equation}\label{1}
S=\frac{1}{2\kappa^2}\int f\left(R,\mathbf{T}^{2}
\right){\sqrt{-g}}d^4x+\int L_{m}\sqrt{-g}d^4x,
\end{equation}
where $\kappa^{2}$, $g$ and $L_{m}$ demonstrate the coupling
constant, determinant of the metric tensor and matter lagrangain,
respectively. This action implies that EMSG has extra degrees of
freedom. Consequently, the possibility of analytic solutions
increases as compared to GR. It is anticipated that some useful
outcomes would be achieved to study the cosmic mysteries in this
gravity due to the matter-dominated era. The action's variation
corresponding to $g_{\alpha\beta}$ yields the equations of motion
\begin{equation}\label{2}
R_{\alpha\beta}f_{R}+g_{\alpha\beta}\Box f_{R}-\nabla
_{\alpha}\nabla_{\beta}f_{R}-\frac{1}{2}g_{\alpha\beta }f
=\kappa^{2}T_{\alpha\beta}-\Theta_{\alpha\beta}f_{ \mathbf{T}^{2}},
\end{equation}
where $\Box= \nabla_{\alpha}\nabla^{\alpha}$, $f\equiv f(R,
\mathbf{T}^{2})$, $f_{\mathbf{T}^{2}}= \frac{\partial f}
{\partial\mathbf{T}^{2}}$, $f_{R}= \frac{\partial f} {\partial R}$,
and
\begin{eqnarray}\label{3}
\Theta_{\alpha\beta} =-2L_{m}\left(T_{\alpha\beta}-\frac{1}{2}g
_{\alpha\beta}T\right)-4\frac{\partial^{2} L_{m}}{\partial
g^{\alpha\beta}
\partial g^{\mu\nu}}T^{\mu\nu}-TT_{\alpha
\beta}+2T_{\alpha}^{\mu}T_{\beta\mu}.
\end{eqnarray}
It is noted that this theory leads to $f(R)$ gravity for
$f(R,T_{\alpha\beta}T^{\alpha\beta})= f(R)$ and reduces to GR when
$f(\mathcal {R},T_{\alpha\beta}T^{\alpha\beta})=R$. In gravitational
physics, the configuration of matter and energy is determined by the
stress-energy tensor and each non-zero component yields dynamical
variables with certain physical attributes.

Here, we take isotropic matter distribution as
\begin{equation}\label{4}
T^{m}_{\alpha\beta}
=\left(\mathrm{\rho}_{m}+\mathrm{p}_{m}\right)\emph{U}_{\alpha}\emph{U}_{\beta}+g_{\alpha\beta}\mathrm{p}_{m},
\end{equation}
where $\emph{U}_{\alpha}$, $\mathrm{p}_{m}$ and $\mathrm{\rho}_{m}$
demonstrate the four velocity, pressure and energy density,
respectively. Manipulating Eq.(\ref{3}), we obtain
\begin{eqnarray}\nonumber
\Theta_{\alpha\beta} =
-\left(3\mathrm{p}_{m}^2+\mathrm{\rho}_{m}^2+4\mathrm
{p}_{m}\mathrm{\rho}_{m}\right)\emph{U}_{\alpha}\emph{U} _{\beta}.
\end{eqnarray}
Rearranging Eq.(\ref{2}), we have
\begin{equation}\label{5}
G_{\alpha\beta} =\kappa^{2}\left(\frac{T_{\alpha\beta}^{c}}{\kappa
^{2}}+\frac{T_{\alpha\beta}^{m}}{f_{R}}\right)=T_{\alpha\beta}^{eff},
\end{equation}
where $G_{\alpha\beta}$ is the Einstein tensor and
$T_{\alpha\beta}^{eff}$ are the additional impacts of EMSG that
include the higher-order curvature terms because of the modification
in curvature part named as correction terms defined as
\begin{equation}\label{6}
T_{\alpha\beta}^{eff}=\frac{1}{2} g_{\alpha\beta}\left(f-Rf_{R}
\right)+\left(\nabla_{\alpha}\nabla_{\beta}-g_{
\alpha\beta}\Box\right)f_{R}-\Theta_{\alpha
\beta}f_{\mathbf{T}^{2}}.
\end{equation}
The $f(R,\mathbf{T}^{2})$ gravity provides non-conserved
stress-energy tensor implying the presence of an extra force which
acts as a non-geodesic motion of particles given by
\begin{equation}\label{7}
\nabla^{\alpha}T^{m}_{\alpha\beta}
=-\frac{1}{2\kappa^2}\Big(f_{\mathbf{T}^{2}}
g_{\alpha\beta}\nabla^{\alpha}\mathbf{T}^{2}
-2\nabla^{\alpha}\left(f_{\mathbf{T}^{2}}
\Theta_{\alpha\beta}\right)\Big).
\end{equation}

In order to study the WH geometry, we consider static spherically
symmetric spacetime as \cite{28}
\begin{equation}\label{8}
ds^{2}=-e^{\lambda\left(\emph{r}\right)}dt^{2}+e^{\vartheta
\left(\emph{r}\right)}dr^{2}+\mathrm{M}\left(\emph{r}
\right)\left(d\theta^{2}+\sin^{2}\theta d\phi^{2}\right),
\end{equation}
where $\mathrm{M}(\emph{r})=\sinh\emph{r}$, $\emph{r}^2$,
$\sin\emph{r}$ for $\emph{K}= -1,0,1$ ($\emph{K}$ defines the
curvature parameter) and $\lim_{\emph{r} \longrightarrow
0}\mathrm{M}(\emph{r})=0$ represents the geodesic deviation equation
\cite{29}. To analyze the WH geometry, we assume
$\mathrm{M}(\emph{r})=\emph{r}^2$ and $e^{\vartheta(\emph{r})}=
\left(1-\frac{b(\emph{r})}{\emph{r}}\right)^{-1}$, where
$\lambda(\emph{r})$ and $b(r)$ define the red-shift and shape
function, respectively. In order to identify the WH throat, the
behavior of $r$ should be non-monotonic as it decreases from
infinity to $\emph{r}_{0}$ (minimum value) and after that it
increases from $\emph{r}_{0}$ to infinity $(\emph{r}>\emph{r}_{0})$
indicating WH throat at $b(\emph{r} _{0})=\emph{r}_{0}$. The
condition $b'(\emph{r}_{0})<1$ must be satisfied to examine the WH
solution at throat, where prime depicts the rate of change
corresponding to radial coordinate. The flaring-out condition
$\frac{b(\emph{r})-\emph{r}b(\emph{r})'}{b(\emph{r})^2}>0$ is the
fundamental feature of WH geometry. For the appearance of
traversable WH, the surface must be independent of horizon as well
as $\lambda(\emph{r})$  must be finite everywhere. The resulting
equations of motion are

\begin{eqnarray}\nonumber
&&e^{\lambda-\vartheta}\left(\frac{M'^{2}}{4M^{2}}+\frac{e^{\vartheta}}{M}-\frac{M''}{M}+\frac{\vartheta'M'}{2M}\right)
=\frac{e^{\lambda}}{f_{R}}\left\{\rho_{m}+\frac{1}{2}\left(Rf_{R}-f\right)
\right.\\\label{8a}&&+\left. e^{-\vartheta}f_{R}''+e^{-\vartheta}
\left(\frac{M'}{M}-\frac{\vartheta'}{2}\right)f_{R}'+\left(3\mathrm{p}_{m}^2+\mathrm{\rho}_{m}^2+4\mathrm
{p}_{m}\mathrm{\rho}_{m}\right)f_{\mathbf{T}^{2}}\right\},
\\\nonumber
&&M\left(\frac{e^{\vartheta}}{M^{2}}-\frac{M'^{2}}{4M^{3}}-\frac{\lambda'M'}{2M^{2}}\right)=
\frac{e^{\vartheta}}{f_{R}}\left\{p_{m}+\frac{1}{2}\left(f-Rf_{R}\right)
\right.\\\label{8b}&&-\left. e^{-\vartheta}
\left(\frac{M'}{M}+\frac{\lambda'}{2}\right)f_{R}'\right\},
\\\nonumber
&&\frac{e^{-2\vartheta}}{4}\left(M'(\lambda'-\vartheta')+2M''+\frac{1}{M}
\left(\lambda'^{2}-\lambda'\vartheta'+2\lambda''\right)-\frac{M'^{2}}{M}\right)
\\\nonumber
&&=\frac{M}{f_{R}}\left\{p_{m}-\frac{1}{2}\left(f-Rf_{R}\right)+\frac{f_{R}'}{e^{\vartheta}}
\left(\frac{M'}{2}-\frac{M'}{M}-\frac{\lambda'-\vartheta'}{2}\right)
\right.\\\label{8c}&&-\left.
\frac{f_{R}''}{M^{2}e^{\vartheta}}\right\}.
\end{eqnarray}

The energy conditions are the key aspects in determining the
physical existence of some cosmological structures. In order to
analyze the physically viable geometry of WH, these conditions must
be violated. To determine the energy conditions, we write down
Raychaudhari equations as
\begin{eqnarray}\nonumber
&&\frac{d\varphi}{d\tau}+\frac{1}{3}\varphi^{2}-\varsigma_{\alpha\beta}
\varsigma^{\alpha\beta}+\upsilon_{\alpha\beta}\upsilon^{\alpha\beta}
+R_{\alpha\beta}k^{\alpha}k^{\beta}=0, \\\nonumber
&&\frac{d\varphi}{d\tau}+\frac{1}{2}\varphi^{2}-\varsigma_{\alpha\beta}
\varsigma^{\alpha\beta}+\upsilon_{\alpha\beta}\upsilon^{\alpha\beta}
+R_{\alpha\beta}l^{\alpha}l^{\beta}=0,
\end{eqnarray}
where $\varphi$, $\varsigma$, $\upsilon$ $k$, $l$ determine the
expansion scalar, shear and rotation tensors, timelike and null
vectors, respectively. These equations are defined for null and
timelike congruences. In GR, these bounds can be categorized into
null $(\mathbb{NEC})$  $(\mathrm{\rho}_{m}+\mathrm{p}_{m}\geq0)$,
weak $(\mathbb {WEC})$  $(\mathrm{\rho}_{m}+\mathrm{p}_{m}\geq0,
\mathrm{\rho}_{m}\geq0)$, strong $(\mathbb{SEC})$
$(\mathrm{\rho}_{m}+3\mathrm{p}_{m}\geq0)$ and dominant
$(\mathbb{DEC})$ $(\mathrm{\rho}_{m}\pm\mathrm{p}_{m}\geq0)$ energy
conditions \cite{30}. The Raychaudhari equation for non-geodesic
congruences as follows
\begin{eqnarray}\nonumber
\frac{d\varphi}{d\tau}+\frac{1}{3}\varphi^{2}-\varsigma_{\alpha\beta}
\varsigma^{\alpha\beta}+\upsilon_{\alpha\beta}\upsilon^{\alpha\beta}
+R_{\alpha\beta}k^{\alpha}k^{\beta}-A=0,
\end{eqnarray}
where $\textit{A}=
\nabla_{\alpha}\left(\emph{U}^{\alpha}\nabla_{\beta}
\emph{U}^{\beta}\right)$ represents the additional impact of
modified gravity named as acceleration term. The purely geometric
nature of Raychaudhari equations implies that
$T^{m}_{\alpha\beta}l^{\alpha}l^{\beta}-A\geq0$ which can be
replaced by $T^{eff}_{\alpha\beta}l^{\alpha}l^{\beta}-A\geq0$.
Consequently, these conditions follow non-geodesic congruences in
curvature-matter coupled gravity expressed as \cite{31}
\begin{eqnarray}\nonumber
&&\mathbb{NEC}:\quad\mathrm{\rho}_{eff}+\mathrm{p}_{eff}-\textit{A}\geq
0,\\\nonumber &&\mathbb{WEC}:\quad\mathrm{\rho}_{eff}-\textit{A}\geq
0, \quad \mathrm {\rho}_{eff}+\mathrm{p}_{eff}-\textit{A}\geq 0,
\\\nonumber
&&\mathbb{SEC}:\quad\mathrm{\rho}_{eff}+\mathrm{p}_{eff}-\textit{A}\geq
0, \quad \mathrm{\rho}_{eff}+3\mathrm{p}_{eff}-\textit{A}\geq 0,
\\\nonumber
&&\mathbb{DEC}:\quad\mathrm{\rho}_{eff}-\textit{A}\geq 0, \quad
\mathrm {\rho}_{eff}\pm \mathrm{p}_{eff}-\textit{A}\geq 0.
\end{eqnarray}
In modified theories, the violation of $(\mathbb{NEC})$ ensures the
presence of physically viable WH. By using Eqs.(\ref{6}), we obtain
\begin{equation}\label{21}
\mathrm{\rho}_{eff}+\mathrm{p}_{eff}-\textit{A}
=\frac{1}{2e^{\vartheta}}\Big(\frac{\lambda'\mathrm{M}'}{\mathrm{M}}
+\frac{\vartheta'\mathrm{M}'}{\mathrm{M}}+\frac{\mathrm{M}'^{2}}
{\mathrm{M}^{2}}-\frac{2\mathrm{M}''}{\mathrm{M}}\Big),
\end{equation}
where the acceleration term is expressed as
\begin{equation}\label{22}
\textit{A}
=\frac{1}{4e^{\vartheta}}\Big(\lambda'^{2}+2\lambda''+4\lambda'\emph{r}
^{-1}\Big)+\frac{\lambda'\left(b-\emph{r}b'\right)}{4r^2}.
\end{equation}

\section{Point-Like Lagrangian}

Here, we formulate point-like Lagrangian corresponding to the action
(\ref{1}) by applying Lagrange multiplier approach as
\begin{equation}\label{9}
S=2\pi^{2}\int \sqrt{-g}\Big\{f-(R-\bar{R})\mu_{1}-
\left(\mathbf{T}^{2}-\mathbf{\bar{T}}^{2}
\right)\mu_{2}+\mathrm{p}_{m}\left(\lambda,
\vartheta,\mathrm{M}\right)\Big\}dr,
\end{equation}
where
\begin{eqnarray}\nonumber
\sqrt{-g}=e^{\frac{\lambda+\vartheta}{2}}\mathrm{M}, \quad
\mathbf{\bar{T}^{2}}=3\mathrm{p}_{m}^{2}+\mathcal{\rho}_{m}^{2},
\quad \mu_{1}=f_{R}, \quad \mu_{2}=f_{\mathbf{T}^{2}},
\\\label{10}
\bar{R}=-\frac{1}{e^{\vartheta}}\Big(\lambda''+\frac
{\lambda'^{2}}{2}+\frac{2\mathrm{M}''}{\mathrm
{M}}+\frac{\lambda'\mathrm{M}'}{\mathrm{M}}-\frac
{\mathrm{M}'^{2}}{2\mathrm{M}^{2}}-\frac{\vartheta'
\mathrm{M}'}{\mathrm{M}}-\frac{\lambda'\vartheta'}
{2}-\frac{2e^{\vartheta}}{\mathrm{M}} \Big).
\end{eqnarray}
We note that the action (\ref{9}) reduces to the action (\ref{1})
for $R-\bar{R}=0$ and $\mathbf{T}^{2}-\mathbf{\bar{T}}^{2}=0$.
Substituting the values from Eq.(\ref{10}) in (\ref{9}) and
eliminating the boundary terms, we have
\begin{eqnarray}\nonumber
&&\mathcal{L}\left(\lambda,\vartheta,\mathrm{M},R,\mathbf{T}^{2}
,\lambda',\vartheta',\mathrm{M}',R',(\mathbf{T}^{2})'\right)=
\mathrm{M}e^{\frac{\lambda+\vartheta}{2}}\Big(f+\mathrm{p}_{m}
-f_{R}\left(R-2\mathrm{M}^{-1}\right)\\\nonumber&&
+f_{\mathbf{T}^{2}}\left(3\mathrm{p}_{m}^{2}+\mathrm{\rho}_{m}^{2}
-\mathbf{T}^{2}\right)\Big)+\mathrm{M}e^{\frac{\lambda-\vartheta}
{2}}\Bigg\{\left(\frac{\lambda'\mathrm{M}'}{\mathrm{M}}+\frac
{\mathrm{M}'^{2}}{2\mathrm{M}^{2}}\right)f_{R}\\\label{11}
&&+\left(\frac{2\mathrm{M}'R'}{\mathrm{M}}+\lambda'R'\right)f_{RR}
+\left(\frac{2\mathrm{M}'(\mathbf{T}^{2})'}{\mathrm{M}}+\lambda'
(\mathbf{T}^{2})'\right)f_{R\mathbf{T}^{2}}\Bigg\}.
\end{eqnarray}
The Euler-Lagrange equations and Hamiltonian of the Lagrangian is
expressed as
\begin{equation}\label{12}
\frac{\partial \mathcal{L}}{\partial q^{i}}-\frac{d}
{dr}\left(\frac{\partial \mathcal{L}}{\partial q^{i'}}\right)=0,
\quad H=q^{i'}\left(\frac{\partial \mathcal{L}} {\partial
q^{i'}}\right)-\mathcal{L}.
\end{equation}
where $q^{i}$ are the generalized coordinates of $n$-dimensional
space. By using Lagrangian (\ref{11}), Eq.(\ref{12}) becomes
\begin{eqnarray}\nonumber
&&f-Rf_{R}+\mathrm{p}_{m}+f_{\mathbf{T}^{2}}\left(3\mathrm{p}_{m}^{2}
+\mathrm{\rho}_{m}^{2}+12\mathrm{p}_{m}\mathrm{p}_{m_{,\lambda}}+4
\mathrm{\rho}\mathrm{\rho}_{m_{,\lambda}}-\mathbf{T}^{2}\right)
+2\mathrm{p}_{m_{,\lambda}}\\\nonumber&&-\frac{1}{e^{\vartheta}}
\Bigg\{\left(\frac{2\mathrm{M}''}{\mathrm{M}}-\frac{\mathrm{M}'^{2}}
{2\mathrm{M}^{2}}-\frac{\vartheta'\mathrm{M}'}{\mathrm{M}}-\frac{2e^{
\vartheta}}{\mathrm{M}}\right)f_{R}+\left(2R''-\vartheta'R'+\frac
{2\mathrm{M}'R'}{\mathrm{M}}\right)f_{RR}\\\nonumber&&+\left(2(\mathbf
{T}^{2})''-\vartheta'(\mathbf{T}^{2})'+\frac{2\mathrm{M}'(\mathbf{T}^{2})'}
{\mathrm{M}}\right)f_{R\mathbf{T}^{2}}+2R'^{2}f_{RRR}+4R'(\mathbf{T}^{2})'f_{RR\mathbf{T}^{2}}\\\label{13}&&
+2((\mathbf{T}^{2})')^{2}f_{R\mathbf{T}^{2}\mathbf{T}^{2}}\Bigg\}=0,\\\nonumber&&
f-Rf_{R}+\mathrm{p}_{m}+f_{\mathbf{T}^{2}}
\left(3\mathrm{p}_{m}^{2}+\mathrm{\rho}_{m}^{2}+12\mathrm{p}_{m}\mathrm{p}_
{m_{,\vartheta}}+4\mathrm{\rho}\mathrm{\rho}_{m_{,\vartheta}}-\mathbf{T}^{2}\right)
\\\nonumber&&+2\mathrm{p}_{m_{,\vartheta}}+\frac{1}{e^{\vartheta}}\Bigg\{\left
(\frac{2e^{\vartheta}}{\mathrm{M}}-\frac{\mathrm{M}'^{2}}{2\mathrm{M}^{2}}-\frac
{\lambda'\mathrm{M}'}{\mathrm{M}}\right)f_{R}-\left(\lambda'R'+\frac{2\mathrm{M}'R'}
{\mathrm{M}}\right)f_{RR}\\\label{14}&&-\left(\lambda'(\mathbf{T}^{2})'+\frac{2\mathrm{M}'
(\mathbf{T}^{2})'}{\mathrm{M}}\right)f_{R\mathbf{T}^{2}}\Bigg\}=0,\\\nonumber&&f-R
f_{R}+f_{\mathbf{T}^{2}}\left(3\mathrm{p}_{m}^{2}+\mathrm{\rho}_{m}^{2}+6\mathrm{M}
\mathrm{p}_{m}\mathrm{p}_{m_{,\mathrm{M}}}+2\mathrm{M}\mathrm{\rho}\mathrm{\rho}_{m_
{,\mathrm{M}}}-\mathbf{T}^{2}\right)\\\nonumber&&-\frac{1}{e^{\vartheta}}\Bigg\{\left
(\lambda''+\frac{\lambda'^{2}}{2}+\frac{\mathrm{M}''}{\mathrm{M}}+\frac{\lambda'\mathrm{M}'}
{2\mathrm{M}}-\frac{\vartheta'\mathrm{M}'}{2\mathrm{M}}-\frac{\lambda'\vartheta'}{2}-\frac
{\mathrm{M}'^{2}}{2\mathrm{M}^{2}}\right)f_{R}\\\nonumber&&+\left(\lambda'R'-\vartheta'R'
+2R''+\frac{\mathrm{M}'R'}{\mathrm{M}}\right)f_{RR}+\left(\lambda'(\mathbf{T}^{2})
'-\vartheta'(\mathbf{T}^{2})\right.\\\nonumber&&\left.+2(\mathbf{T}^{2})''+\frac
{\mathrm{M}'(\mathbf{T}^{2})'}{\mathrm{M}}\right)f_{R\mathbf{T}^{2}}+2R'^{2}f_{RRR}
+4R'(\mathbf{T}^{2})'f_{RR\mathbf{T}^{2}}\\\label{15}&&
+2((\mathbf{T}^{2})')^{2}f_{R\mathbf{T}^{2}\mathbf{T}^{2}}
-\mathrm{p}_{m}e^{\vartheta} -\mathrm{M}\mathrm{p}_{m_{,\mathrm{M}}}
e^{\vartheta}\Bigg\}=0,
\\\nonumber&&\left(\lambda''+\frac{\lambda'^{2}}{2}+\frac{2\mathrm{M}''}
{\mathrm{M}}+\frac{\lambda'\mathrm{M}'}{\mathrm{M}}-\frac{\vartheta'\mathrm{M}'}{\mathrm{M}}-\frac
{\lambda'\vartheta'}{2}-\frac{\mathrm{M}'^{2}}{2\mathrm{M}^{2}}\right)f_{RR}\\\label{16}&&
+e^{\vartheta}\bigg\{\left(R-2\mathrm{M}^{-1}\right)f_{RR}-\left(3\mathrm{p}_{m}^{2}+\mathrm{\rho}_{m}^{2}
-\mathbf{T}^{2}\right)f_{R\mathbf{T}^{2}}\bigg\}=0,\\\nonumber&&\left(\lambda''+\frac{\lambda'^{2}}
{2}+\frac{2\mathrm{M}''}{\mathrm{M}}+\frac{\lambda'\mathrm{M}'}{\mathrm{M}}-\frac{\vartheta'\mathrm{M}'}
{\mathrm{M}}-\frac{\lambda'\vartheta'}{2}-\frac{\mathrm{M}'^{2}}{2\mathrm{M}^{2}}\right)f_{R\mathbf{T}^{2}}
\\\label{17}&&+e^{\vartheta}\bigg\{\left(R-2\mathrm{M}^{-1}\right)f_{R\mathbf{T}^{2}}-\left(3\mathrm{p}_{m}
^{2}+\mathrm{\rho}_{m}^{2}-\mathbf{T}^{2}\right)f_{\mathbf{T}^{2}\mathbf{T}^{2}}\bigg\}=0.
\end{eqnarray}
The variation of energy function corresponding to Lagrangian
(\ref{11}) yields
\begin{equation}\label{20}
e^{\vartheta(r)}
=\frac{\left(\frac{\mathrm{M}'^{2}}{2\mathrm{M}^{2}}+\frac{\lambda'
\mathrm{M}'}{\mathrm{M}}\right)f_{R}+\left(\lambda'+\frac{2\mathrm
{M}'}{\mathrm{M}}\right)\Big(R'f_{RR}+(\mathbf{T}^{2})'f_{\mathbf{T}
^{2}\mathbf{T}^{2}}\Big)}{\Big(f-Rf_{R}+\left(3\mathrm{p}_{m}^{2}+
\mathrm{\rho}_{m}^{2}-\mathbf{T}^{2}\right)f_{\mathbf{T}^{2}}+\mathrm
{p}_{m}+\frac{2f_{R}}{\mathrm{M}}\Big)}.
\end{equation}

\section{Noether Symmetry Approach}

Noether symmetries are used to discuss the solutions of dynamical
configuration and also their existence provides some viable
conditions of cosmological models according to current observations
\cite{32}. In particular, the Noether symmetry strategy is also used
to probe the nature of mysterious energy \cite{33}-\cite{36}. The
main incentive comes from different laws of conservation that are
consequences of some type of symmetry that exists in a system. The
conservation laws are the major aspects in the study of different
physical phenomena and every continuous symmetry yields the
conservation law as indicated by the Noether theorem. This theorem
is important as it offers a relation among symmetries and conserved
entities of the system. The EMSG is a non-conserved theory but we
attain conserved quantities in the framework of the Noether symmetry
technique. These are useful to derive exact or numeric solutions to
examine the mysterious universe. To examine the presence of Noether
symmetry with corresponding conserved quantity, we consider
\begin{equation}\label{23}
Y= \varrho(\lambda,\vartheta,\mathrm{M},R,
\mathbf{T}^{2})\frac{\partial}{\partial \emph{r}}
+\zeta(\lambda,\vartheta,\mathrm{M},R,
\mathbf{T}^{2})^{i}\frac{\partial}{\partial q^{i}}, \quad i=
1,2,3,4,5.
\end{equation}
where $\varrho$ and $\zeta$ are unknown coefficients of the vector
field. The Lagrangian must satisfy the invariance condition to enure
the presence of Noether symmetries. Accordingly, $Y$ plays a role of
symmetry generator which establishes the conserved quantities. The
invariance condition is determined as
\begin{equation}\label{24}
Y^{[1]}\mathcal{L}+(D\varrho)\mathcal{L}= D\psi,
\end{equation}
where $Y^{[1]}$ defines the prolongation of first order, $D$
represents the total rate of change and $\psi$ is the boundary
term,. Further, it is determined as
\begin{equation}\label{25}
Y^{[1]}= Y+{\zeta^{i}}'\frac{\partial}{\partial {q^{i}}'},
~~~D=\frac{\partial}{\partial \emph{r}}+{q^{i}}'\frac{\partial}
{\partial {q^{i}}},
\end{equation}
here ${\zeta^{i}}'= D{\zeta^{i}}'-{q^{i}}'D\varrho$.

The conserved quantities associated with symmetry generators are
expressed as
\begin{equation}\label{26}
I= -\varrho H+ \zeta^{i}\frac{\partial \mathcal{L}}{\partial
q^{i}}-\psi.
\end{equation}
This is the most important part of Noether symmetries that plays a
key role to derive physically viable solutions. By considering
Eq.(\ref{24}) and comparing the coefficients
${\lambda'}^2\mathrm{M}',~ \lambda' \vartheta'\mathrm{M}',~
\lambda'{\mathrm{M}'}^2,~ \lambda'{R'}^2$ and
$\lambda'(\mathbf{T}^2)'$, we obtain
\begin{eqnarray}\label{27}
\varrho_{,\lambda}f_{R}=0, \quad \varrho_{,\vartheta}f_{R}=0, \quad
\varrho_{,\mathrm{M}}f_{R}=0, \quad \varrho_{,R}f_{RR}=0, \quad
\varrho_{,\mathbf{T}^{2}}f_{R\mathbf{T}^{2}}=0.
\end{eqnarray}
This shows that either $\varrho_{,\lambda}, ~\varrho_{,\vartheta},~
\varrho_{,\mathrm{M}},~ \varrho_{,R},~ \varrho_{,\mathbf{T}^{2}}=0$,
or $f_{R},~ f_{RR}, ~f_{R\mathbf{T}^{2}}=0$. For the second choice,
we get a trivial solution. So, for the non-trivial solution $f_{R},~
f_{RR},~ f_{R\mathbf{T}^{2}}\neq0$ and equating the remaining
coefficients, we have the following system of equations
\begin{eqnarray}\label{28}
&&\psi_{,\vartheta}=0, \quad \varrho_{,\lambda}=0, \quad
\varrho_{,\vartheta}=0, \quad \varrho_{,\mathrm{M}}=0, \quad
\varrho_{,R}=0, \quad \varrho_{,\mathbf{T} ^{2}}=0.\\\label{29}
&&\mathrm{M}\zeta^{1}_{,\vartheta}f_{RR}+2\zeta^{3}_{,\vartheta}f_{RR}=0,
\\\label{30}
&&\mathrm{M}\zeta^{1}_{,R}f_{RR}+2\zeta^{3}_{,R}f_{RR}=0,
\\\label{31}
&&\mathrm{M}\zeta^{1}_{,\vartheta}f_{R\mathbf{T}^{2}}+2\zeta^{3}
_{,\vartheta}f_{R\mathbf{T}^{2}}=0,
\\\label{32}
&&\mathrm{M}\zeta^{1}_{,\mathbf{T}^{2}}f_{R\mathbf{T}^{2}}+2\zeta^{3}
_{,\mathbf{T}^{2}}f_{R\mathbf{T}^{2}}=0,
\\\label{33}
&&\zeta^{3}_{,\vartheta}f_{R}+\mathrm{M}\zeta^{4}_{,\vartheta}f_{RR}
+\mathrm{M}\zeta^{5}_{,\vartheta}f_{R\mathbf{T}^{2}}=0,
\\\label{34}
&&\mathrm{M}\zeta^{1}_{,\emph{r}}f_{RR}+2\zeta^{3}_{,\emph{r}}f_{RR}
-e^{\frac{\vartheta-\lambda}{2}}\psi_{,R}=0,
\\\label{35}
&&\zeta^{3}_{,\lambda}f_{R}+\mathrm{M}\zeta^{4}_{,\lambda}f_{RR}+\mathrm{M}
\zeta^{5}_{,\lambda}f_{R\mathbf{T}^{2}}=0,
\\\label{36}
&&\mathrm{M}\zeta^{1}_{,\emph{r}}f_{R\mathbf{T}^{2}}+2\zeta^{3}_{,\emph{r}}
f_{R\mathbf{T}^{2}}-e^{\frac{\vartheta-\lambda}{2}}\psi_{,\mathbf{T}^{2}}=0,
\\\label{37}
&&\zeta^{1}_{,\vartheta}f_{R}+\zeta^{3}_{,\vartheta}\mathrm{M}^{-1}f_{R}
+2\zeta^{4}_{,\vartheta}f_{RR}+2\zeta^{5}_{,\vartheta}f_{R\mathbf{T}^{2}}=0,
\\\label{38}
&&\zeta^{3}_{,\emph{r}}f_{R}+\mathrm{M}\zeta^{5}_{,\emph{r}}f_{RR}+\mathrm{M}
\zeta^{5}_{,\emph{r}}f_{R\mathbf{T}^{2}}-e^{\frac{\vartheta-\lambda}{2}}
\psi_{,\lambda}=0,
\\\label{39}
&&\mathrm{M}\zeta^{1}_{,\mathbf{T}^{2}}f_{RR}+2\zeta^{3}_{,\mathbf{T}^{2}}f_{RR}+
\mathrm{M}\zeta^{1}_{,R}f_{R\mathbf{T}^{2}}+2\zeta^{3}_{,R}f_{R\mathbf{T}^{2}}=0,
\\\label{40}
&&\zeta^{1}_{,\emph{r}}f_{R}+\zeta^{3}_{,\emph{r}}\mathrm{M}^{-1}f_{R}+2\zeta^
{4}_{,\emph{r}}f_{RR}+2\zeta^{5}_{,\emph{r}}f_{R\mathbf{T}^{2}}-e^{\frac{\vartheta
-\lambda}{2}}\psi_{,\mathrm{M}}=0, \\\nonumber
&&\left(\zeta^{1}-\zeta^{2}-2\mathrm{M}^{-1}\zeta^{3}+4\mathrm{M}\zeta^{1}_{,\mathrm{M}}
+4\zeta^{3}_{,\mathrm{M}} -2\varrho_{,\emph{r}}\right)f_{R}
\\\label{41}
&&+\left(2\zeta^{4}+8\mathrm{M}\zeta^{4}_{,\mathrm{M}}\right)f_{RR}+\left(2\zeta
^{5}+8\mathrm{M}\zeta^{5}_{,\mathrm{M}}\right)f_{R\mathbf{T}^{2}}=0,
\\\nonumber
&&\left(2\zeta^{4}+2\mathrm{M}\zeta^{4}_{,\mathrm{M}}+4\zeta^{4}_{,\lambda}\right)
f_{RR}+\left(2\zeta^{5}+2\mathrm{M}\zeta^{5}_{,\mathrm{M}}+4\zeta^{5}_{,\lambda}\right)
f_{R\mathbf{T}^{2}}
\\\label{42}
&&+\left(\zeta^{1}-\zeta^{2}+2\zeta^{1}_{,\lambda}+2\mathrm{M}^{-1}\zeta^{3}_{,\lambda}
+2\zeta^{3}_{,\mathrm{M}}-2\varrho_{,\emph{r}}\right)f_{R}=0,
\\\nonumber
&&\left(\zeta^{1}-\zeta^{2}+\mathrm{M}\zeta^{1}_{,\mathrm{M}}+2\zeta^{3}_{,\mathrm{M}}
+2\zeta^{4}_{,R}-2\varrho_{,r}\right)f_{RR}+2\zeta^{4}f_{RRR}
\\\label{43}
&&+\left(\zeta^{1}_{,R}+\mathrm{M}^{-1}\zeta^{3}_{,R}\right)f_{R}+2\zeta^{5}f_{RR\mathbf{T}^{2}}
+2\zeta^{5}_{,R}f_{R\mathbf{T}^{2}}=0,
\\\nonumber
&&\left(\zeta^{1}-\zeta^{2}+\mathrm{M}\zeta^{1}_{,\mathrm{M}}+2\zeta^{3}_{,\mathrm{M}}+2\zeta^{5}
_{,\mathbf{T}^{2}}-2\varrho_{,\emph{r}}\right)f_{R\mathbf{T}^{2}}+2\zeta^{4}f_{RR\mathbf{T}^{2}}
\\\label{44}
&&+\left(\zeta^{1}_{,\mathbf{T}^{2}}+\mathrm{M}^{-1}\zeta^{3}_{,\mathbf{T}^{2}}\right)f_{R}
+2\zeta^{5}f_{R\mathbf{T}^{2}\mathbf{T}^{2}}+2\zeta^{4}_{,\mathbf{T}^{2}}f_{RR}=0,
\\\nonumber
&&\left(\mathrm{M}\zeta^{1}-\mathrm{M}\zeta^{2}+2\zeta^{3}+2\mathrm{M}\zeta^{1}_{,\lambda}+4\zeta^{3}
_{,\lambda}+2\mathrm{M}\zeta^{4}_{,R}-2\mathrm{M}\varrho_{,\emph{r}}\right)f_{RR}
\\\label{45}
&&+2\zeta^{3}_{,R}f_{R}+2\mathrm{M}\zeta^{4}f_{RRR}+2\mathrm{M}\zeta^{5}f_{RR\mathbf{T}^{2}}+2\zeta^{5}
_{,R}f_{R\mathbf{T}^{2}}=0,
\\\nonumber
&&\left(\mathrm{M}\zeta^{1}-\mathrm{M}\zeta^{2}+2\zeta^{3}+2\mathrm{M}\zeta^{1}_{,\lambda}+4\zeta^{3}
_{,\lambda}+2\mathrm{M}\zeta^{4}_{,R}-2\mathrm{M}\varrho_{,r}\right)f_{R\mathbf{T}^{2}}
\\\label{46}
&&+2\zeta^{3}_{,\mathbf{T}^{2}}f_{R}+2\mathrm{M}\zeta^{4}f_{RR\mathbf{T}^{2}}+2\mathrm{M}
\zeta^{5}f_{R\mathbf{T}^{2}\mathbf{T}^{2}}+2\zeta^{5}_{,\mathbf{T}^{2}}f_{RR}=0,
\\\nonumber
&&e^{\frac{\lambda+\vartheta}{2}}\mathrm{M}\bigg\{\left(f-R
f_{R}+\mathrm{p}_{m}+f_{\mathbf{T}^{2}}\left(3\mathrm{p}_{m}^{2}+\mathrm{\rho}_{m}^{2}
-\mathbf{T}^{2}\right)+2\mathrm{M}^{-1}f_{R}\right)
\\\nonumber
&&\times\left(\frac{\zeta^{1}+\zeta^{2}}{2}+\varrho_{,r}\right)
+\zeta^{1}\left(f_{\mathbf{T}^{2}}\left(6\mathrm{p}_{m}\mathrm{p}_{m_{,\lambda}}+2\mathrm
{\rho}\mathrm{\rho}_{m_{,\lambda}}\right)+\mathrm{p}_{m_{,\lambda}}\right)
\\\nonumber
&&+\zeta^{2}\left(f_{\mathbf{T}^{2}}\left(6\mathrm{p}_{m}\mathrm{p}_{m_{,\vartheta}}+2\mathrm
{\rho}\mathrm{\rho}_{m_{,\vartheta}}\right)+\mathrm{p}_{m_{,\vartheta}}\right)
+\zeta^{3}\left(f_{\mathbf{T}^{2}}\left(6\mathrm{p}_{m}\mathrm{p}_{m_{,\mathrm{M}}}+2\mathrm
{\rho} \right.\right.\\\nonumber
&&\times\left.\left.\mathrm{\rho}_{m_{,\mathrm{M}}}\right)+\mathrm{p}_{m_{,\mathrm{M}}}\right)
+\frac{\zeta^{3}}{\mathrm{M}}\left(f-R
f_{R}+\mathrm{p}_{m}+f_{\mathbf{T}^{2}}\left(3\mathrm{p}_{m}
^{2}+\mathrm{\rho}_{m}^{2}-\mathbf{T}^{2}\right)\right)
\\\nonumber
&&-\zeta^{4}\left(f_{RR}\left(R-2\mathrm{M}^{-1}\right)+f_{R\mathbf{T}^{2}}\left(3\mathrm{p}_{m}^{2}
+\mathrm{\rho}_{m}^{2}-\mathbf{T}^{2}\right)\right)-\zeta^{5}\left(f_{R\mathbf{T}^{2}}\left(R
\right.\right.\\\label{47}
&&\left.\left.-2\mathrm{M}^{-1}\right)+f_{\mathbf{T}^{2}\mathbf{T}^{2}}\left(3\mathrm{p}_{m}^{2}
+\mathrm{\rho}_{m}^{2}-\mathbf{T}^{2}\right)\right)\bigg\}-\psi_{,\emph{r}}=0.
\end{eqnarray}

Noether symmetry approach reduces the system's complexity and helps
in determining the exact solutions. Therefore, the analysis of
viable and traversable WH solutions through this strategy would
provide fascinating results. However, the above system is highly
nonlinear and complicated because of the multivariate functions and
their derivatives. It is difficult to find a non-trivial solution
without considering any specific EMSG model. In the following, we
take minimal model as \cite{36a}
\begin{itemize}
\item $f(R,\mathbf{T}^{2})=R+\eta(\mathbf{T}^{2})^{n}$.
\end{itemize}
where $\eta$ is a constant. We consider $\eta=1$ for the sake of
simplicity. In order to make resemblance of this model with the
standard $\Lambda$CDM model, we add cosmological constant in this
model and redefine as
\begin{equation}\label{48}
f(R,\mathbf{T}^{2})=R+\Lambda(\mathbf{T}^{2})+(\mathbf{T}^{2})^{n}.
\end{equation}
The simultaneous solutions of Eqs.(\ref{28})-(\ref{46}) yield
\begin{eqnarray}\nonumber
&&\zeta^{2}=-\frac{2\emph{c}_{2}\emph{c}_{5}}{\emph{r}^{2}}, \quad
\varrho= \emph{c}_{1}-\frac {\emph{c}_{2}\emph{c}_{5}}{\emph{r}},
\quad \zeta^{1}=\zeta^{3}=\zeta^{4}=\zeta^{5}=0,\\\label{49}
&&\Lambda(\mathbf{T}^{2})=-(\mathbf{T}^{2})^{n}
+\emph{c}_{3}\mathbf{T}^{2}+\emph{c}_{4}, \quad \psi=
\emph{c}_{5}\emph{r},
\end{eqnarray}
where $\emph{c}_{i}$ represent the arbitrary constants.

It is noteworthy to examine perfect matter as it describes the exact
matter configuration of different astrophysical objects. The cosmic
matter configuration can also be examined by dust matter only when a
negligible amount of radiation is present. In the following, we
analyze the presence of viable traversable WH and derive exact
solutions of $f(R,\mathbf{T}^{2})$ gravity model for dust and
non-dust matter distributions.

\subsection{Dust Case}

For dust matter distribution, Eq.(\ref{4}) reduces to
\begin{equation}\label{50}
T^{m}_{\alpha\beta}=\mathcal{\rho}_{m}\emph{U}_{\alpha}\emph{U}_{\beta}.
\end{equation}
Using Eq.(\ref{50}) in (\ref{47}), we obtain
\begin{eqnarray}\label{51}
\mathcal{\rho}_{m} =
\sqrt{\frac{e^\frac{-\lambda-\vartheta}{2}}{2\emph{c}_{2}
\emph{c}_{3}}}, \quad  f(R,\mathbf{T}^{2})=
R+2\emph{c}_{3}\mathbf{T}^{2}+\emph{c}_{4}.
\end{eqnarray}
The symmetry generators and corresponding conserved quantities
become
\begin{eqnarray}\nonumber
Y_{1}&=&\frac{\partial}{\partial \emph{r}}, \quad
Y_{2}=-\frac{2\emph{c}_{2}}{\emph{r}} \frac{\partial}{\partial
\emph{r}}-\frac{2\emph{c}_{2}}{\emph{r}^{2}}\frac{\partial}
{\partial \vartheta},\\\nonumber
I_{1}&=&2e^{\frac{\lambda-\vartheta}{2}}\bigg\{1+\lambda'\emph{r}-\left(1+\frac
{\emph{c}_{4}\emph{r}^{2}}{2}+\frac{\emph{r}^{2}e^{\frac{-\lambda-\vartheta}{2}}}
{2\emph{c}_{2}}\right)e^{\vartheta}\bigg\},\\\nonumber
I_{2}&=&\emph{r}-\frac{2\emph{c}_{2}e^{\frac{\lambda-\vartheta}{2}}}{\emph{r}}\bigg
\{1+\lambda'\emph{r}-\left(1+\frac{\emph{c}_{4}\emph{r}^{2}}{2}+\frac{\emph{r}^{2}
e^{\frac{-\lambda-\vartheta}{2}}}{2\emph{c}_{2}}\right)e^{\vartheta}\bigg\}.
\end{eqnarray}
Substituting Eq.(\ref{51}) in (\ref{20}), we have
\begin{equation}\label{52}
e^{\vartheta(\emph{r})} =
\frac{1+\lambda'\emph{r}}{1+\frac{\emph{r}^{2}\emph{c}_{4}}{2}
+\frac{\emph{r}^{2}e^{\frac{-\lambda-\vartheta}{2}}}{2\emph{c}_{2}}}.
\end{equation}
We consider both constant as well as variable red-shift function
$(\lambda(\emph{r})=h$, $\lambda(\emph{r})= -h/\emph{r};~ h>0)$
\cite{37} to study the structure and existence of a physically
viable WH via energy bounds and shape function. In the following, we
manipulate Eq.(\ref{52}) for both values of the red-shift function.

\subsubsection*{Case I: $\lambda(r)=h$}

Inserting this value in (\ref{52}), we have
\begin{equation}\label{53}
\vartheta(\emph{r}) = 2\ln
\bigg\{-\frac{\emph{r}^{2}e^{\frac{-h}{2}}+\sqrt
{\left(e^{\frac{-h}{2}}\right)^{2}\emph{r}^{4}+8\emph{r}
^{2}\emph{c}_{2}^{2}\emph{c}_{4}+16\emph{c}_{2}^{2}}}{2\emph{c}_{2}
\left(\emph{r}^{2}\emph{c}_{4}+2\right)}\bigg\}.
\end{equation}
The associated shape function is
\begin{eqnarray}\nonumber
b(\emph{r}) &=&
\Big\{2\emph{r}^{3}\left(\emph{r}^{2}e^{-h}-4\emph{c}_{2}
^{2}\emph{c}_{4}+e^{\frac{-h}{2}}\sqrt{\emph{r}^{4}e^{-h}+8
\emph{r}^{2}\emph{c}_{2}^{2}\emph{c}_{4}+16\emph{c}_{2}^{2}}\right)
\\\label{54}
&-&2\emph{r}^{2}\emph{c}_{2}^{2}\emph{c}_{4}^{2}\Big\}\Big\{\left
(\emph{r}^{2}e^{\frac{-h}{2}}+\sqrt{\emph{r}^{4}e^{-h}+8\emph
{r}^{2}\emph{c}_{2}^{2}\emph{c}_{4}+16\emph{c}_{2}^{2}}\right)^{2}\Big\}
^{-1}.
\end{eqnarray}
The energy density for dust matter becomes
\begin{equation}\label{55}
\rho_{m}=\sqrt{\frac{e^{\frac{-h}{2}-\ln \bigg\{-\frac{\emph{r}^{2}
e^{\frac{-h}{2}}+\sqrt{\left(e^{\frac{-h}{2}}\right)^{2}\emph
{r}^{4}+8\emph{r}^{2}\emph{c}_{2}^{2}\emph{c}_{4}+16\emph{c}_{2}^{2}}}
{2\emph{c}_{2}\left(\emph{r}^{2}\emph{c}_{4}+2\right)}\bigg\}}}{2\emph
{c}_{2}\emph{c}_{3}}}.
\end{equation}
By using a graphical representation, we examine the geometry of WH.
In Figure \textbf{1}, the upper left plot implies that the action of
shape function increases positively with $b(\emph{r})<r$ whereas the
right plot is asymptotically flat. The left plot in the below panel
determines the throat of WH at $\emph{r}_{0}=0.01$ and the
associated right plot shows
$\frac{db\left(\emph{r}_{0}\right)}{d\emph{r}}<1$. To analyze the
existence of traversable WH, we substitute Eq.(\ref{54}) in
(\ref{21}) as
\begin{eqnarray}\nonumber
\mathrm{\rho}_{eff}+\mathrm{p}_{eff}-\textit{A}
&=&\Big\{e^{\frac{-h}{2}}\sqrt{\emph{r}^{4}e^{-h}+8\emph
{r}^{2}\emph{c}_{2}^{2}\emph{c}_{4}+16\emph{c}_{2}^{2}}+\emph{r}^{2}
e^{-h}
\\\nonumber
&-&2\emph{r}^{2}\emph{c}_{2}^{2}\emph{c}_{4}^{2}-4\emph{c}_{2}^{2}
\emph{c}_{4}\Big\}\left(64\emph{c}_{2}^2+32\emph{r}^2\emph{c}_{2}^2
\emph{c}_{4}\right)
\\\nonumber
&\times&\Big\{\left(\emph{r}^{4}e^{-h}+8\emph{r}^{2}\emph
{c}_{2}^{2}\emph{c}_{4}+16\emph{c}_{2}^{2}\right)^{\frac{1}{2}}+
\left(\emph{r}^{2}e^{\frac{-h}{2}} \right.\\\label{56}
&+&\left.\left(8\emph{r}^{2}\emph{c}_{2}^{2}\emph{c}_{4}+16\emph
{c}^{2}_{2}+r^{4}e^{-h}\right)^{\frac{1}{2}}\right)^{3}\Big\}^{-1}.
\end{eqnarray}
Figure \textbf{2} describes that the behavior of energy density is
positively increasing whereas the effective matter variables are
negatively increasing ($\rho_{m}-\textit{A}>0$ and $\mathrm
{\rho}_{eff}+\mathrm{p}_{eff}-\textit{A}<0$). This inequality shows
that matter variables violate $\mathbb{NEC}$ which ensures the
presence of physically realistic traversable WH.
\begin{figure}
\epsfig{file=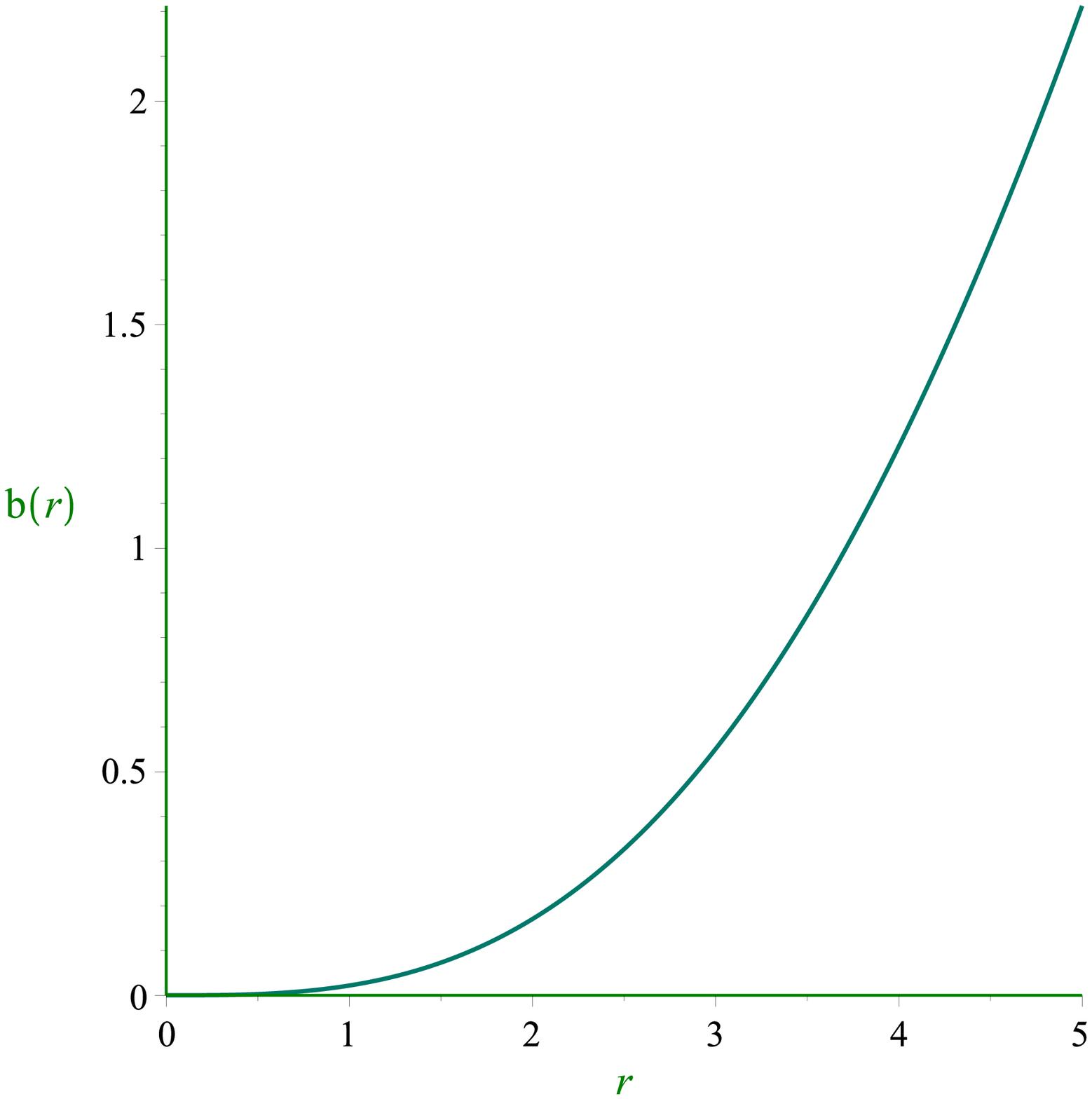,width=.5\linewidth}
\epsfig{file=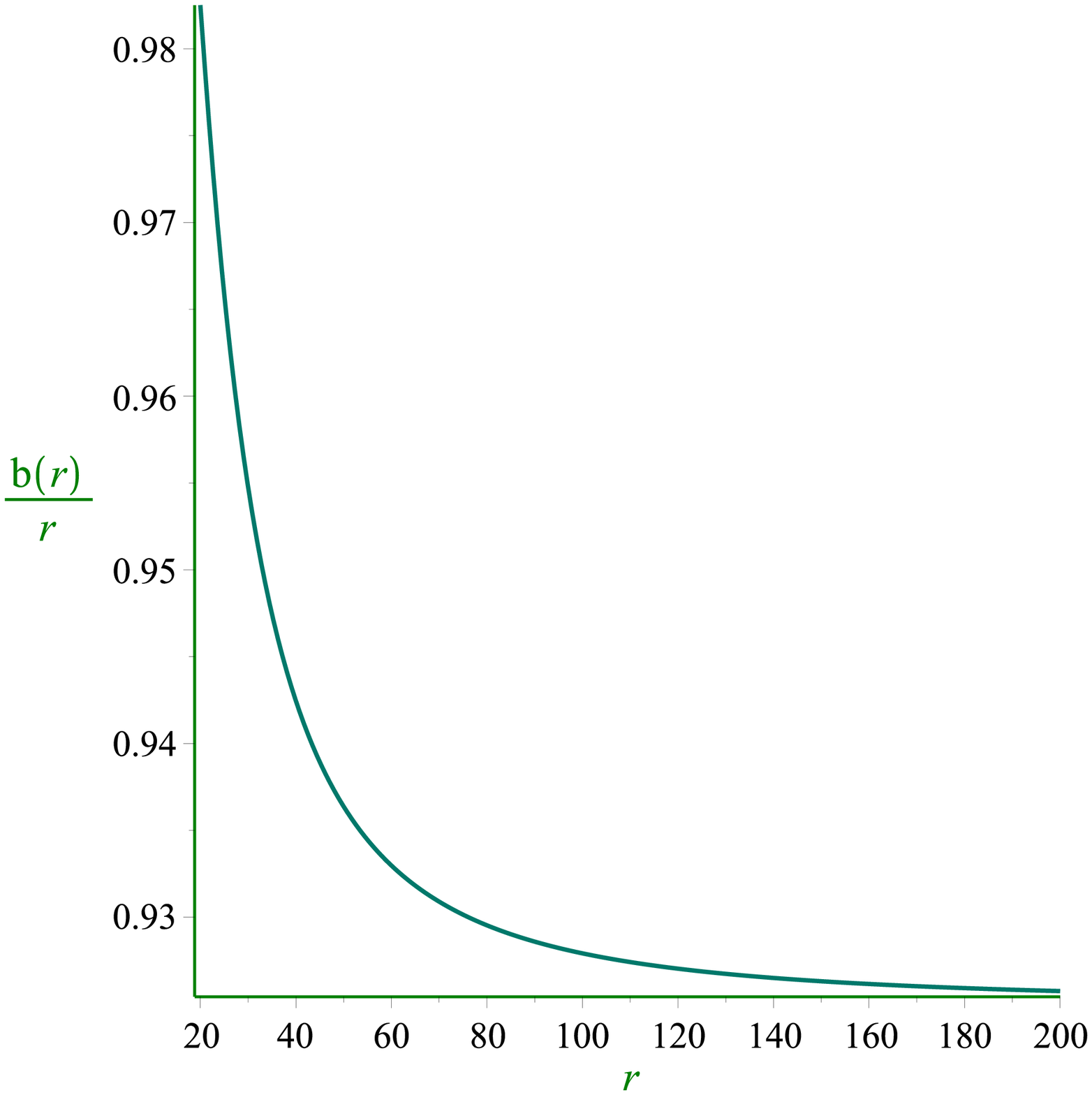,width=.5\linewidth}
\epsfig{file=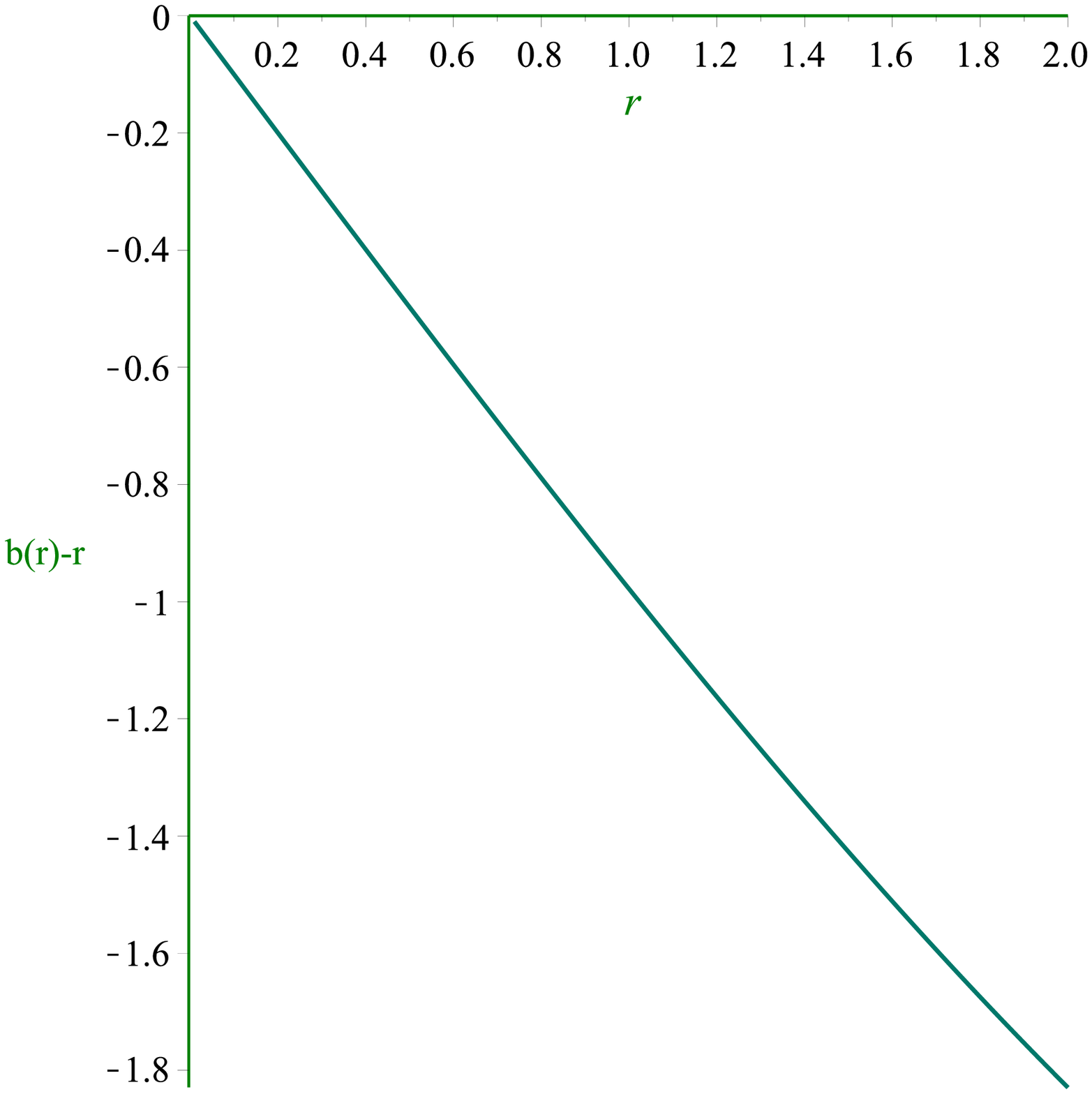,width=.5\linewidth}
\epsfig{file=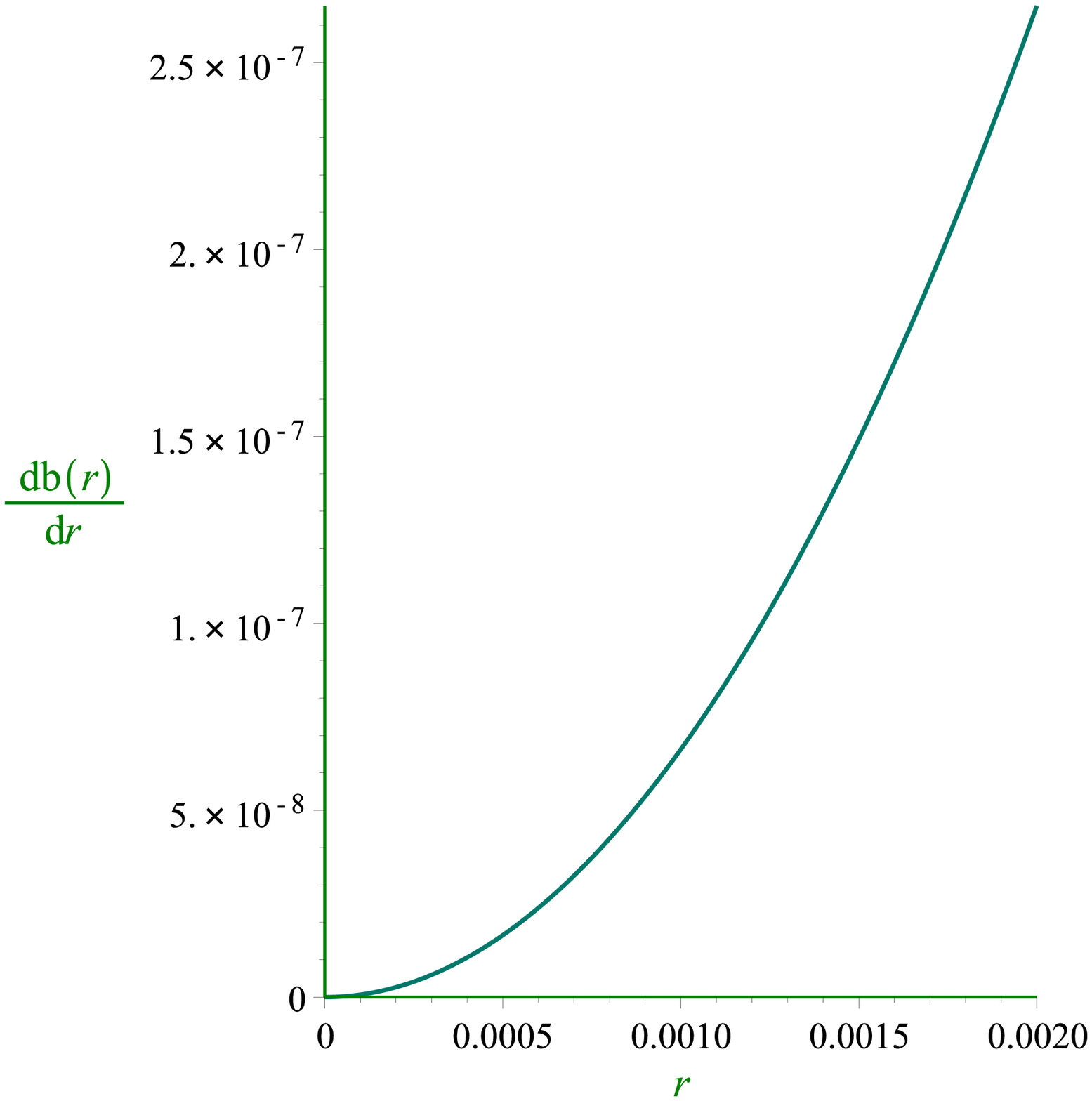,width=.5\linewidth} \caption{Graphs of
$b(\emph{r})$, $\frac{b(\emph{r})}{\emph{r}}$,
$b(\emph{r})-\emph{r}$ and $\frac{db(\emph{r})}{\emph{r}}$
corresponding to $\emph{r}$ for $\emph{c}_{2}$=30,
$\emph{c}_{4}$=-0.0095 and h=-0.08.}
\end{figure}
\begin{figure}
\epsfig{file=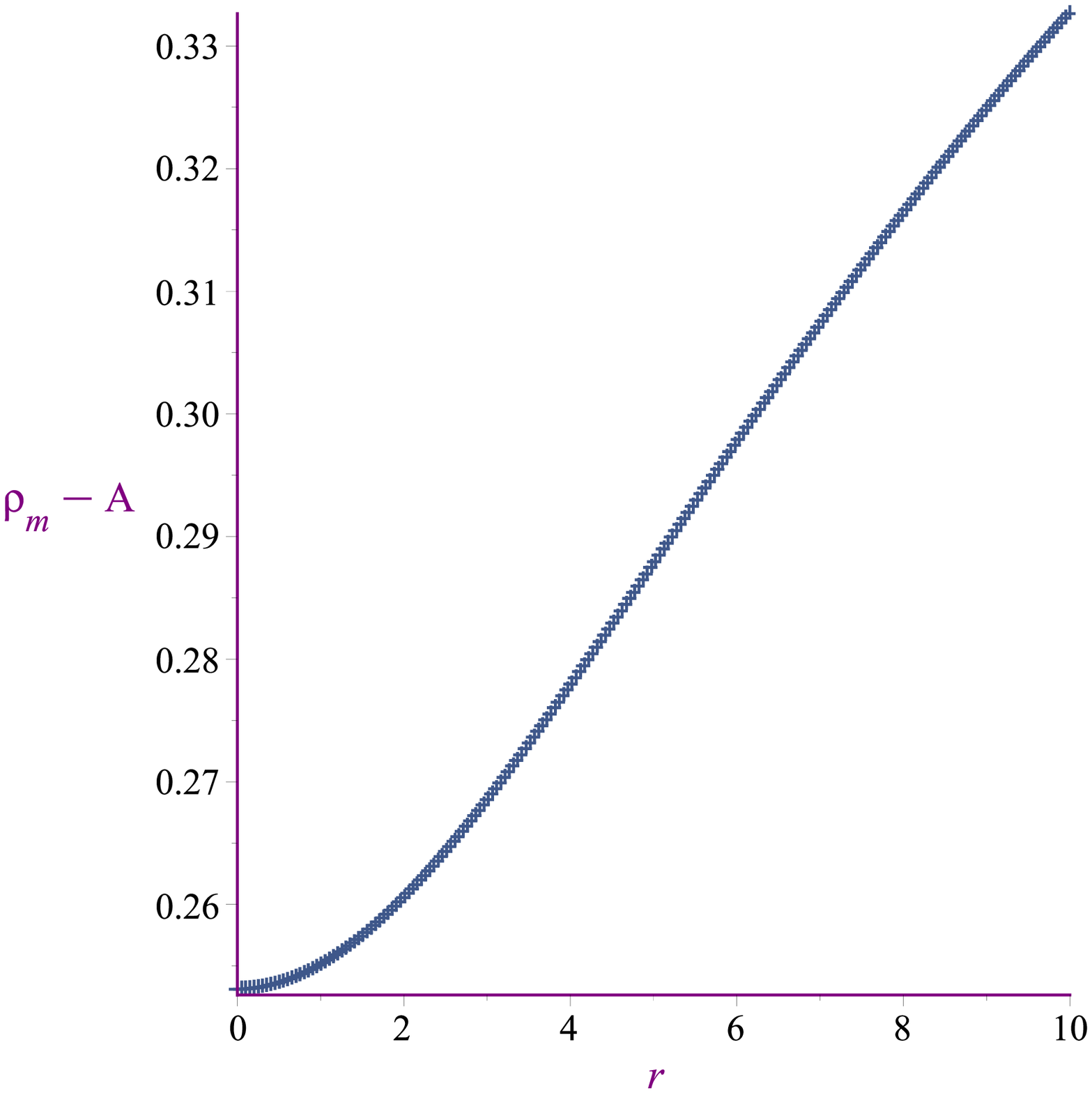,width=.5\linewidth}
\epsfig{file=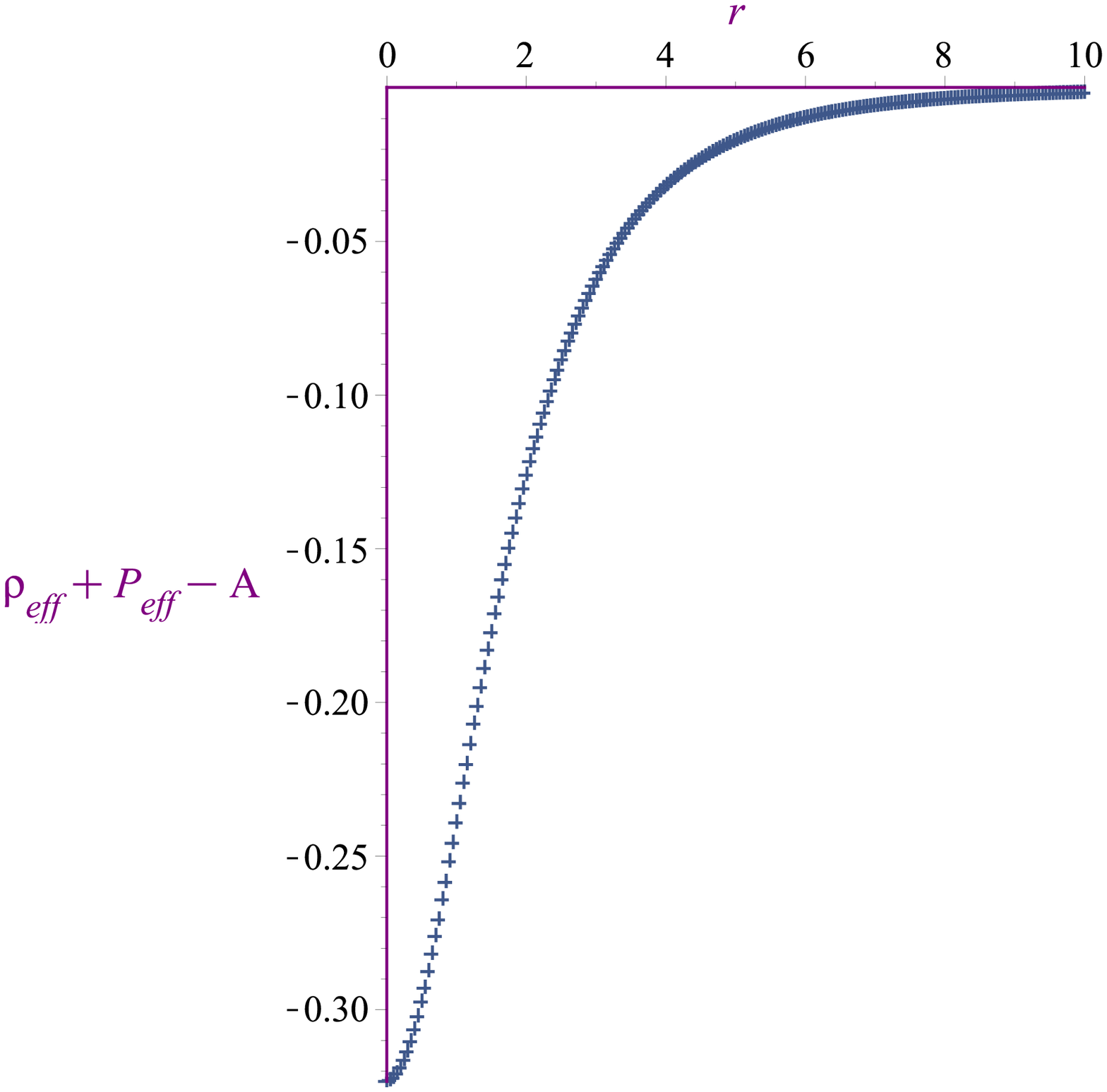,width=.5\linewidth} \caption{Graphs of
$\rho_{m}-\textit{A}$ and $\mathrm{\rho}_{eff}+\mathrm{p}_{eff}
-\textit{A}$ versus $\emph{r}$}
\end{figure}

\subsubsection*{Case II: $\lambda(r)=-\frac{h}{\emph{r}}$}

Here, Eq.(\ref{52}) yields
\begin{eqnarray}\nonumber
\vartheta(r) &=& 2\ln \left\{-\left(\left(\emph{r}^{6}e^{\frac{h}
{\emph{r}}}+8\emph{r}^{4}\emph{c}_{2}^{2}\emph{c}_{4}+8\emph
{r}^{3}h\emph{c}_{2}^{2}\emph{c}_{4}+16\emph{r}^2\emph{c}_{2}^{2}\right.
\right.\right.\\\label{57}&+&\left.\left.\left.16\emph{r}
h\emph{c}_{2}^{2}\right)^{\frac{1}{2}}+r^{3}e^{\frac{h}{2\emph
{r}}}\right)\left(2\emph{c}_{2}\left(\emph{r}^{2}\emph{c}_{4}+2
\right)\right)^{-1}\right\}.
\end{eqnarray}
The associated shape function becomes
\begin{eqnarray}\nonumber
b(r)&=&2\emph{r}^2\left(\sqrt{\emph{r}^{6}e^{\frac{h}{\emph{r}}}
+8\emph{r}^{4}\emph{c}_{2}^{2}\emph{c}_{4}+8\emph{r}^{3}h\emph{c}_{2}^{2}
\emph{c}_{4}+16\emph{r}^2\emph{c}_{2}^{2}+16\emph{r}h\emph{c}_{2}^{2}}
\right.\\\nonumber&\times&\left.
\emph{r}e^{\frac{h}{2\emph{r}}}+\emph{r}^{5}e^{\frac{h}{\emph{r}}}
+4h\emph{r}^{2}\emph{c}_{2}^{2}\emph{c}_{4}-4\emph{r}^{3}\emph{c}_{2}^{2}
\emph{c}_{4}-2\emph{r}^{5}\emph{c}_{2}^{2}\emph{c}_{4}^{2}+8h\emph{c}_{2}^{2}
\right)
\\\nonumber&\times&
\Big\{\left(\left(\emph{r}^{6}e^{\frac{h}{\emph{r}}}+8\emph{r}^{4}
\emph{c}_{2}^{2}\emph{c}_{4}+8\emph{r}^{3}h\emph{c}_{2}^{2}\emph{c}_{4}+16\emph
{r}^2\emph{c}_{2}^{2}+16\emph{r}h\emph{c}_{2}^{2}\right)^{\frac{1}{2}}
\right.\\\nonumber&+&\left.
\emph{r}^{3}e^{\frac{h}{2\emph{r}}}\right)^{2}\Big\}^{-1}.
\end{eqnarray}
The corresponding energy density takes the form
\begin{eqnarray}\nonumber
\mathcal{\rho}_{m} = \sqrt{\frac{e^{\frac{-\lambda}{2}-\ln
\left\{-\left (\emph{r}^{6}e^{\frac{h}{\emph{r}}}+8\emph{r}^{4}\emph
{c}_{2}^{2}\emph{c}_{4}+8\emph{r}^{3}h\emph{c}_{2}^{2}\emph{c}_{4}
+16\emph{r}^2\emph{c}_{2}^{2}+16\emph{r}h\emph{c}_{2}^{2}\right)
^{\frac{1}{2}}-\emph{r}^{3}e^{\frac{h}{2\emph{r}}}\right\}}}
{4\emph{c}_{2}^2\emph{c}_{3}\left(2+\emph{r}^{2}\emph{c}_{4}\right)}}.
\end{eqnarray}
Figure \textbf{3} indicates that the shape function maintains its
positivity and the structure of WH is obtained asymptotically flat.
The left graph in the lower panel exhibits the throat of WH at
$\emph{r}_{0}=0.4$ and the associated right graph implies that
$\frac{db\left(\emph{r}_{0}\right)}{d\emph{r}}<1$. For the existence
of physically viable WH, we substitute the value of $\lambda
(\emph{r})$ and $\vartheta(\emph{r})$ in Eq.(\ref{21}), it gives
\begin{eqnarray}\nonumber
\mathrm{\rho}_{eff}+\mathrm{p}_{eff}-\textit{A} &=&
\left(64\emph{c}_{2}^2+32\emph{r}^2\emph{c}_{2}^2\emph{c}_{4}\right)\left(4h^2\emph{r}^{2}
\emph{c}_{2}^{2}\emph{c}_{4}-4h\emph{r}^{3}\emph{c}_{2}^{2}\emph{c}_{4}
\right.\\\nonumber&-&\left.
4\emph{r}^{4}\emph{c}_{2}^{2}\emph{c}_{4}+\emph{r}^{4}h^2\emph{c}_{2}^{2}\emph{c}_{4}
^{2}-2\emph{r}^{6}\emph{c}_{2}^{2}\emph{c}_{4}^{2}-2h\emph{r}^{5}\emph{c}_{2}^{2}\emph{c}_{4}^{2}
\right.\\\nonumber&+&\left.
4h^2\emph{c}_{2}^2+\emph{r}^{6}e^{\frac{h}{\emph{r}}}\right)+e^{\frac{h}{2\emph{r}
}}\emph{r}^{3}\left(8\emph{r}^{4}\emph{c}_{2}^{2}\emph{c}_{4}+8\emph{r}^{3}h\emph{c}_{2}^{2}\emph{c}_{4}
\right.\\\nonumber&+&\left.
\emph{r}^{6}e^{\frac{h}{\emph{r}}}+16\emph{r}^2\emph{c}_{2}^{2}+16\emph{r}h\emph{c}_{2}
^{2}\right)^{\frac{1}{2}}\Big\{\left(\left(r^{6}e^{\frac{h}{r}}+16r^2\emph{c}_{2}^{2}
\right.\right.\\\nonumber&+&\left.
\left.8\emph{r}^{4}\emph{c}_{2}^{2}\emph{c}_{4}+8\emph{r}^{3}h\emph{c}_{2}^{2}\emph{c}_{4}+16\emph{r}
h\emph{c}_{2}^{2}\right)^{\frac{1}{2}}+r^{3}e^{\frac{h}{2\emph{r}}}\right)^{3}
\\\nonumber&\times&
\left(8\emph{r}^{4}\emph{c}_{2}^{2}\emph{c}_{4}+8\emph{r}^{3}h\emph{c}_{2}^{2}\emph{c}_{4}+\emph{r}^{6}
e^{\frac{h}{\emph{r}}}+16\emph{r}^2\emph{c}_{2}^{2}
\right.\\\nonumber&+&\left.
16\emph{r}h\emph{c}_{2}^{2}\right)^{\frac{1}{2}}\Big\}^{-1}.
\end{eqnarray}
Figure \textbf{4} implies that $\rho_{m}-\textit{A}\geq0$ and
$\mathrm{\rho}_{eff}+\mathrm{p}_{eff}-\textit{A}\leq0$. This
inequality assures the presence of a viable traversable wormhole.
\begin{figure}
\epsfig{file=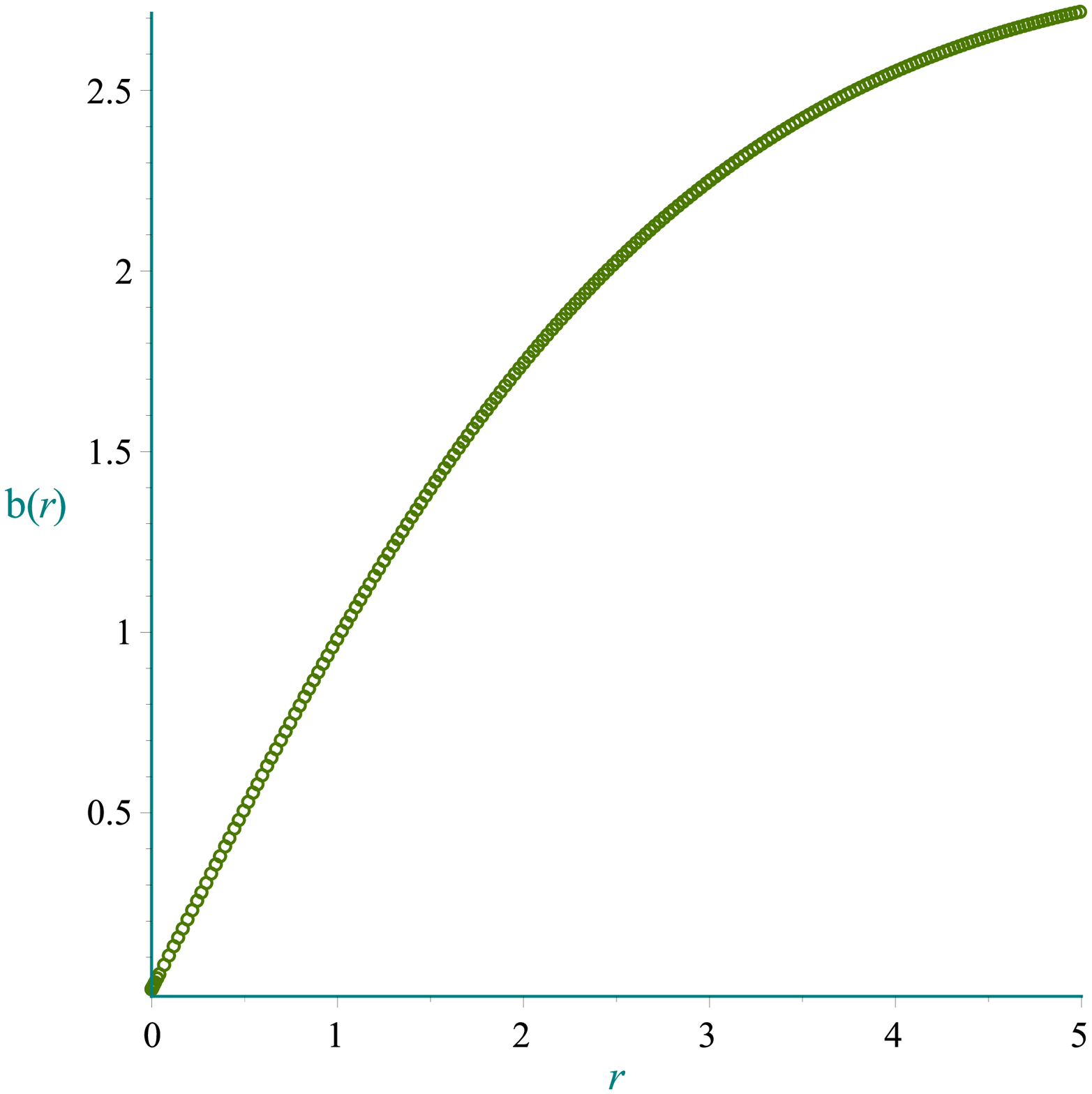,width=.5\linewidth}
\epsfig{file=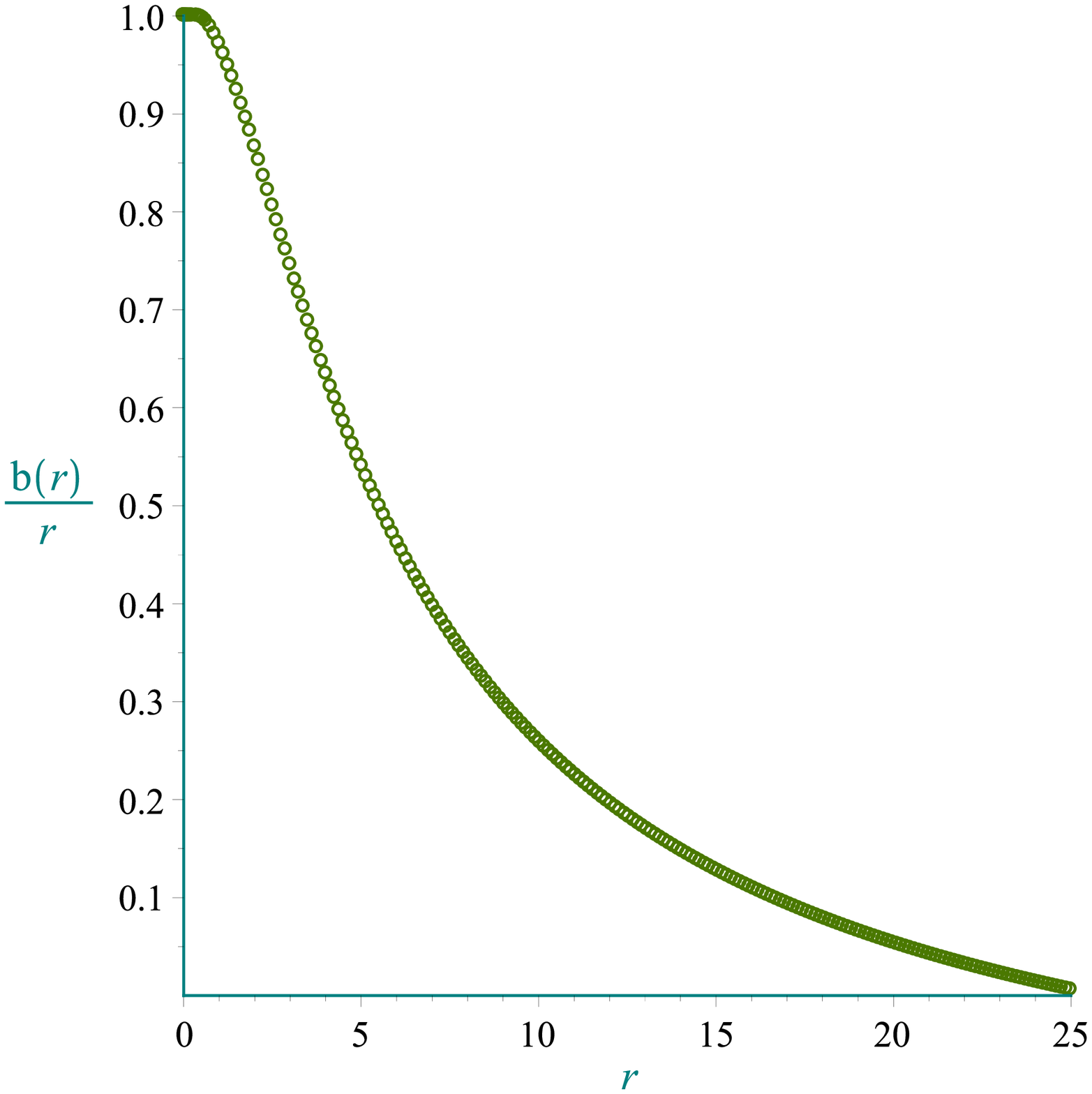,width=.5\linewidth}
\epsfig{file=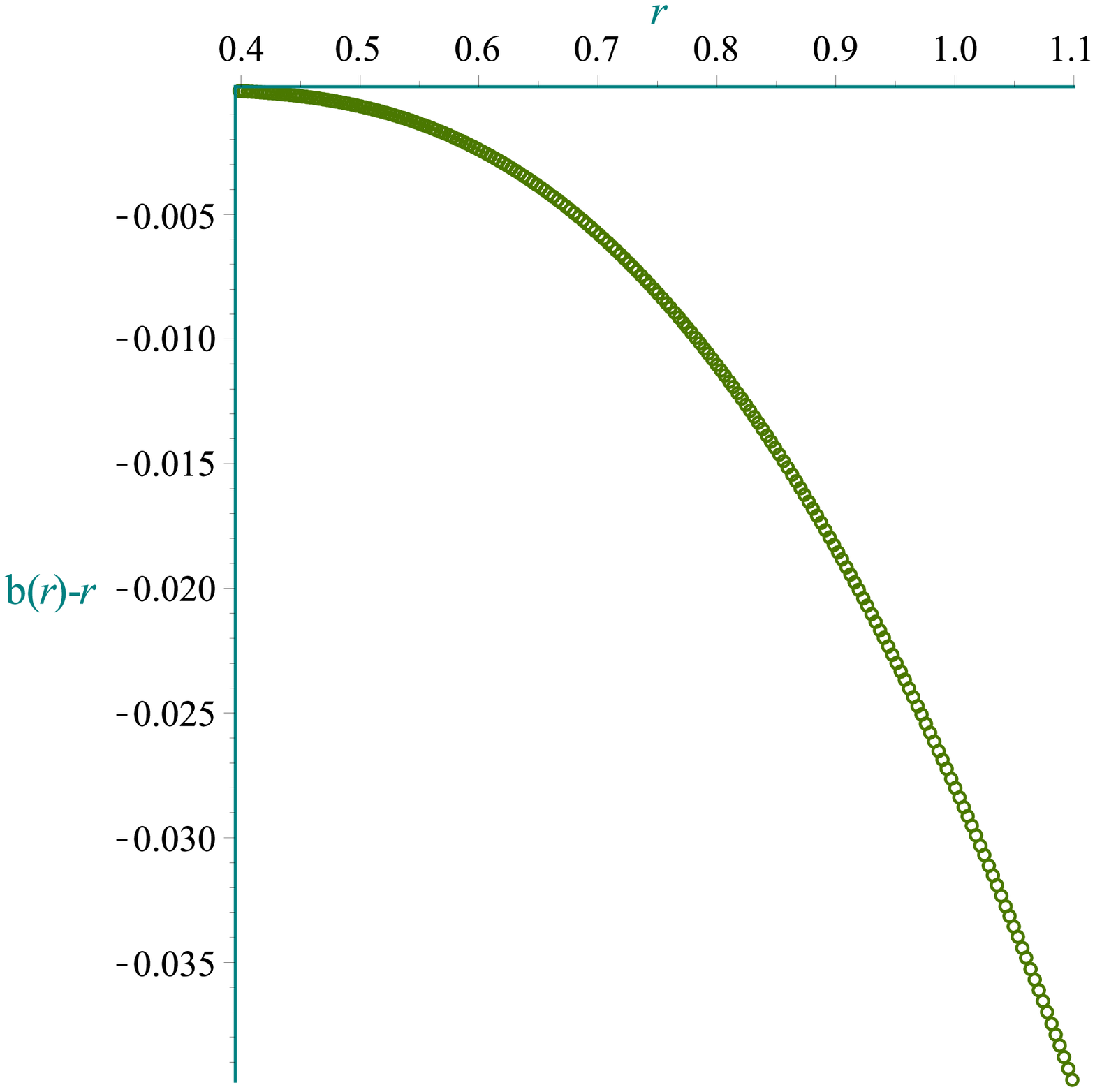,width=.5\linewidth}
\epsfig{file=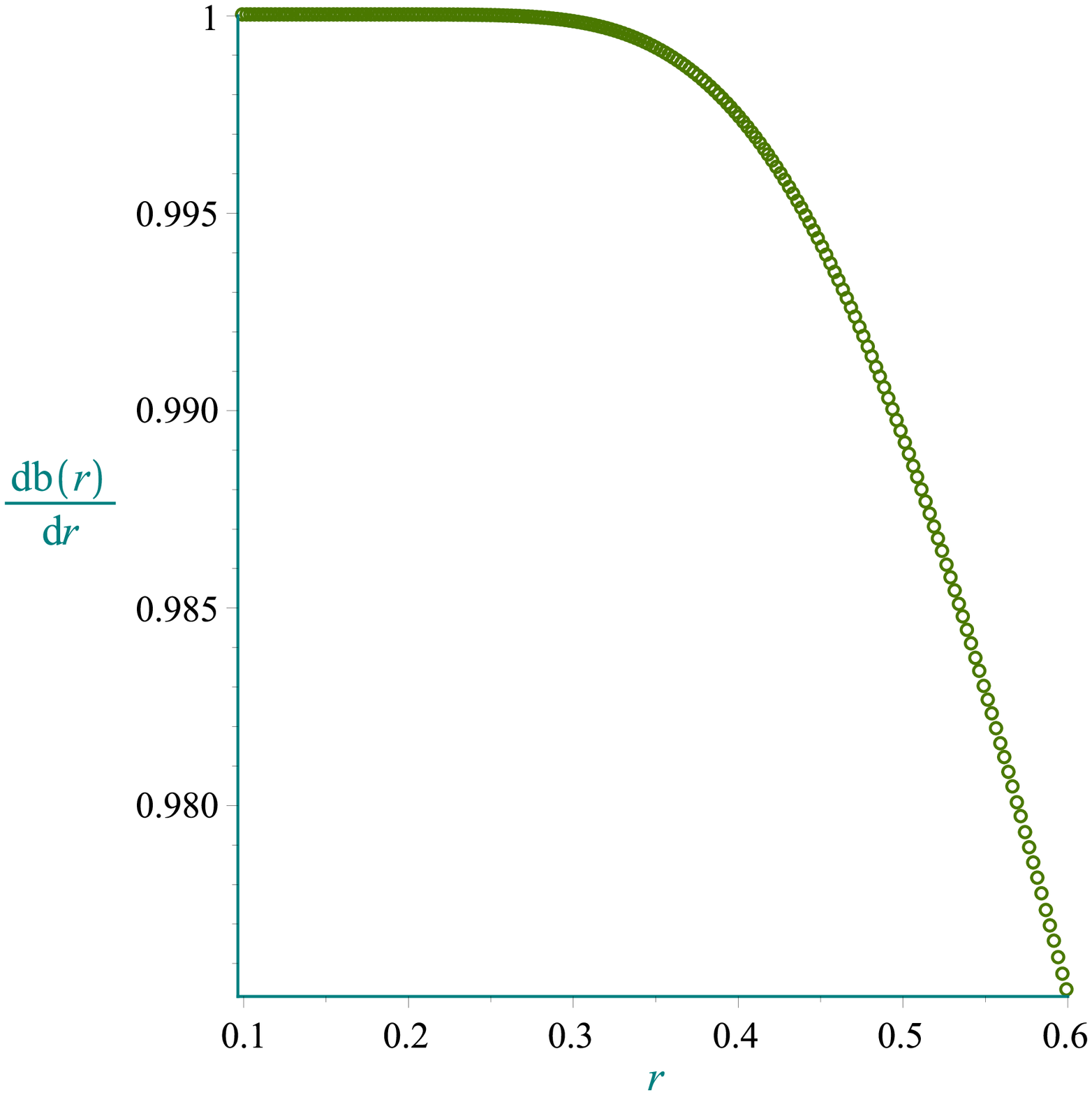,width=.5\linewidth} \caption{Graphs of
$b(\emph{r}),~\frac{b(\emph{r})}{\emph{r}},~b(\emph{r})-\emph{r}$
and $\frac{db(\emph{r})}{\emph{r}}$ corresponding to $\emph{r}$ for
$\emph{c}_{2}$=0.5=$\emph{c}_{3}$, $\emph{c}_{4}$=2.2 and h=4.9.}
\end{figure}
\begin{figure}
\epsfig{file=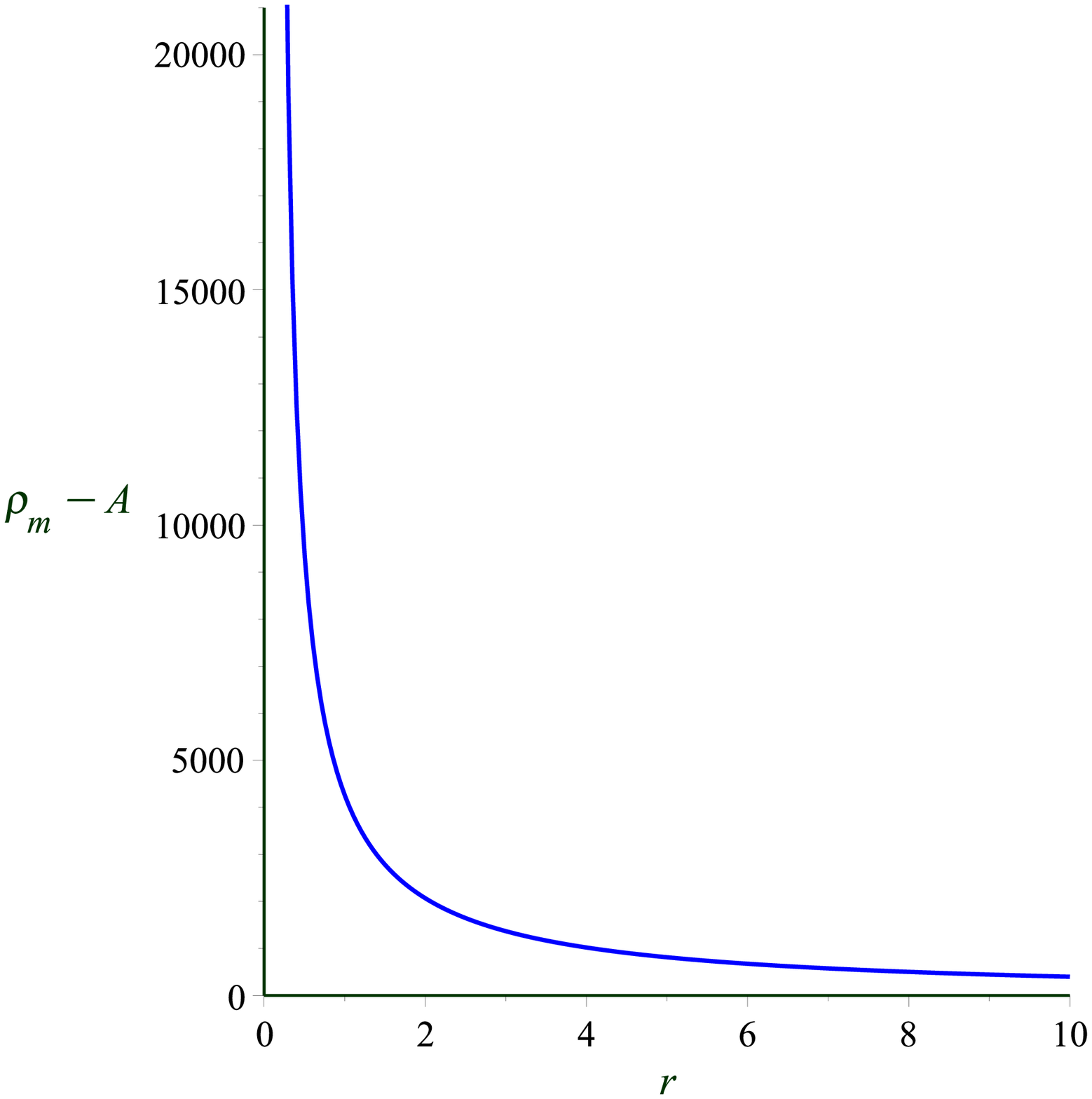,width=.5\linewidth}
\epsfig{file=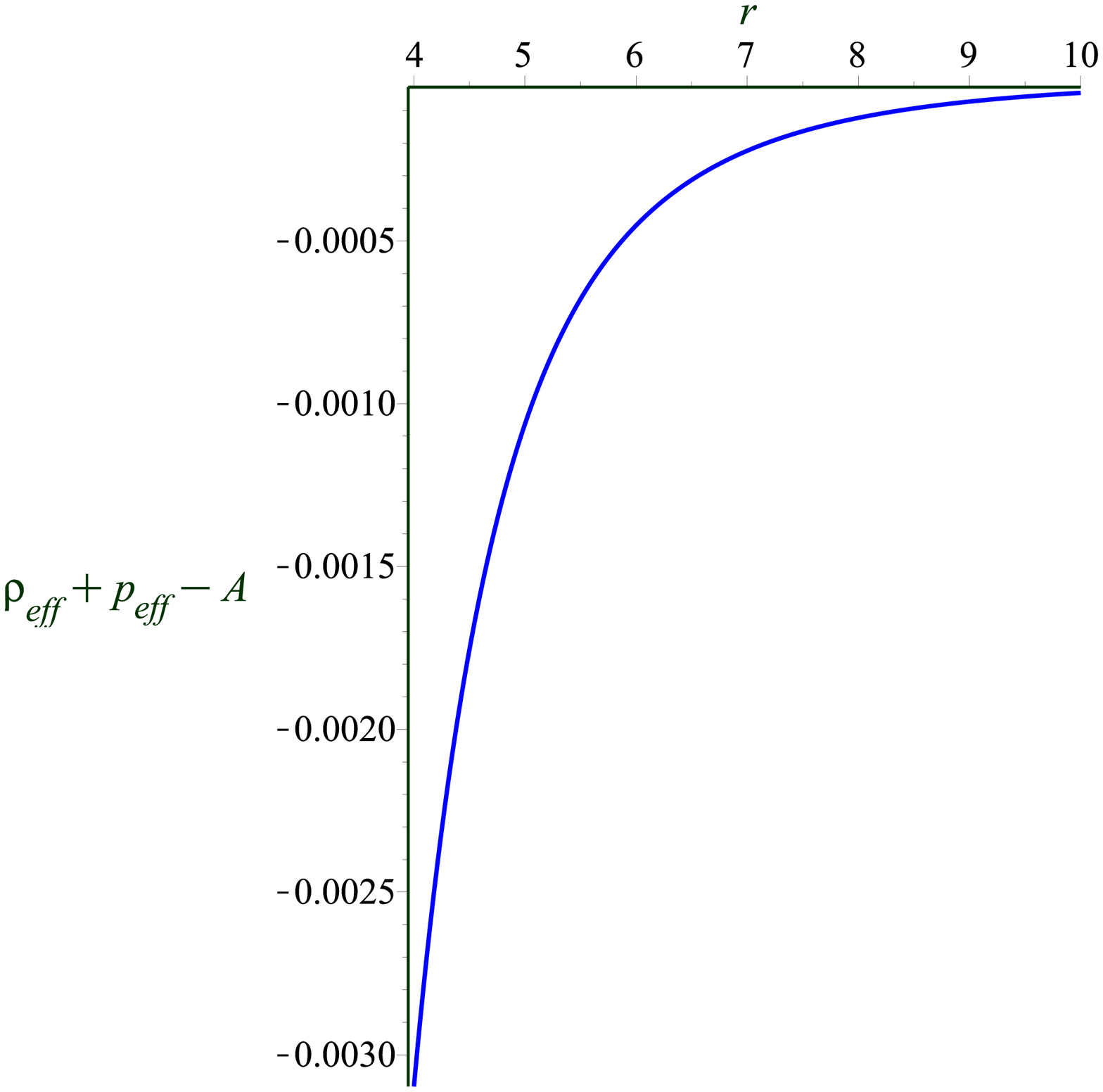,width=.5\linewidth} \caption{Graphs of
$\rho_{m}-\textit{A}$ and
$\mathrm{\rho}_{eff}+\mathrm{p}_{eff}-\textit{A}$ versus $\emph{r}$}
\end{figure}

\subsection{Non-Dust Case}

In the presence of radiations, this case well explains the cosmic
matter configuration. Therefore, we take into account a specific
correlation between matter variables such that
$\mathrm{p}_{m}(\lambda,\vartheta,\mathrm{M})=\omega\mathrm{\rho}_{m}(\lambda,
\vartheta,\mathrm{M})$($\omega$ represents the equation of state
parameter) and manipulate
 Eq.(\ref{47}) which gives
\begin{equation}\label{58}
\mathrm{\rho}_{m} =
\frac{-\emph{c}_{2}\omega+\sqrt{\emph{c}_{2}^2\omega^2+4\emph{c}_{2}
\emph{c}_{3}e^\frac{-\lambda-\vartheta}{2}+12\emph{c}_{2}\emph{c}_{3}
\omega^2e^\frac{-\lambda-\vartheta}{2}}}{2\emph{c}_{2}\emph{c}_{3}
\left(3\omega^2+1\right)}.
\end{equation}
In this case, the generators of Noether symmetry are the same as for
the dust case while the associated conserved quantities are given as
\begin{eqnarray}\nonumber
I_{1} &=&
e^{\frac{\lambda-\vartheta}{2}}\emph{r}^{2}\bigg\{2\left(\frac{1+\lambda'\emph{r}}
{\emph{r}^2}\right)-e^{\vartheta}\left(\emph{c}_{4}+\frac{2}{\emph{r}^2}+\left(\emph{c}_{3}\left(3\omega^2
+1\right)\right) \right.\\\nonumber&\times&\left.
\left(\frac{-\emph{c}_{2}\omega+\sqrt{\emph{c}_{2}^2\omega^2+4\emph{c}_{2}\emph{c}_{3}e^\frac{-\lambda
-\vartheta}{2}+12\emph{c}_{2}\emph{c}_{3}\omega^2e^\frac{-\lambda-\vartheta}{2}}}{2\emph{c}_{2}\emph{c}_{3}
\left(3\omega^2+1\right)}\right)^{2} \right.\\\nonumber&+&\left.
\omega\frac{-\emph{c}_{2}\omega+\sqrt{\emph{c}_{2}^2\omega^2+4\emph{c}_{2}\emph{c}_{3}e^\frac{-\lambda
-\vartheta}{2}+12\emph{c}_{2}\emph{c}_{3}\omega^2e^\frac{-\lambda-\vartheta}{2}}}{2\emph{c}_{2}
\emph{c}_{3}\left(3\omega^2+1\right)}\right) \bigg\}
\\\nonumber
I_{2} &=&
\emph{r}-\emph{c}_{2}\emph{r}e^{\frac{\lambda-\vartheta}{2}}\bigg\{\frac{2\lambda'}
{\emph{r}}+\frac{2}{\emph{r}^2}-e^{\vartheta}\left(\emph{c}_{4}+\frac{2}{\emph{r}^2}+\left
(\emph{c}_{3}\left(3\omega^2+1\right)\right)
\right.\\\nonumber&\times&\left.
\left(\frac{-\emph{c}_{2}\omega+\sqrt{\emph{c}_{2}^2\omega^2+4\emph{c}_{2}\emph{c}_{3}e^\frac{-\lambda
-\vartheta}{2}+12\emph{c}_{2}\emph{c}_{3}\omega^2e^\frac{-\lambda-\vartheta}{2}}}{2\emph{c}_{2}\emph
{c}_{3}\left(3\omega^2+1\right)}\right)^{2}
\right.\\\nonumber&+&\left.
\omega\frac{-\emph{c}_{2}\omega+\sqrt{\emph{c}_{2}^2\omega^2+4\emph{c}_{2}\emph{c}_{3}e^\frac{-\lambda-
\vartheta}{2}+12\emph{c}_{2}\emph{c}_{3}\omega^2e^\frac{-\lambda-\vartheta}{2}}}{2\emph{c}_{2}\emph{c}_{3}
\left(3\omega^2+1\right)}\right) \bigg\}
\end{eqnarray}
Substituting the values of matter variables from the equation of
state and using Eq.(\ref{58}) in (\ref{20}), we obtain
\begin{equation}\label{59}
e^{\vartheta(\emph{r})}=
\frac{2+2\lambda'\emph{r}}{\emph{r}^{2}\left(\emph{c}_{4}
+\frac{2}{\emph{r}^2}+\frac{e^{\frac{-\lambda-\vartheta}{2}}}{\emph{c}_{2}}
\right)}.
\end{equation}
We study the structure and existence of viable WH for the same
red-shift functions as discussed for the dust case.

\subsubsection*{Case I: $\lambda(r)=h$}

In this case, Eq.(\ref{59}) reduces to
\begin{equation}\label{60}
\vartheta(\emph{r}) = 2\ln
\bigg\{\frac{-\emph{r}^{2}e^{\frac{-h}{2}}
+\sqrt{\left(e^{\frac{-h}{2}}\right)^{2}\emph{r}^{4}
+8\emph{r}^{2}\emph{c}_{2}^{2}\emph{c}_{4}+16\emph{c}_{2}^{2}}}
{2\emph{c}_{2}\left(r^{2}\emph{c}_{4}+2\right)}\bigg\}.
\end{equation}
The corresponding shape function becomes
\begin{eqnarray}\nonumber
b(\emph{r}) &=&
\bigg\{2\emph{r}^{3}\left(\emph{r}^{2}e^{-h}-4\emph{c}_{2}^{2}
\emph{c}_{4}-e^{\frac{-h}{2}}\sqrt{\emph{r}^{4}e^{-h}+8\emph{r}^{2}
\emph{c}_{2}^{2}\emph{c}_{4}+16\emph{c}_{2}^{2}}\right)
\\\nonumber
&-&2\emph{r}^{2}\emph{c}_{2}^{2}\emph{c}_{4}^{2}\bigg\}\bigg\{\left(\emph{r}^{2}e^{\frac{-h}{2}}
-\sqrt{\emph{r}^{4}e^{-h}+8\emph{r}^{2}\emph{c}_{2}^{2}\emph{c}_{4}+16\emph{c}_{2}^{2}}\right)^{2}\bigg\}^{-1}.
\end{eqnarray}
Inserting Eq.(\ref{60}) in (\ref{58}), we have
\begin{eqnarray}\nonumber
\mathrm{\rho}_{m} &=&
\bigg\{-\emph{c}_{2}\omega+\bigg\{4\emph{c}_{2}\emph{c}_{3}e^{\frac{-\lambda}
{2}-\ln
\big(\frac{-\emph{r}^{2}e^{\frac{-h}{2}}+\sqrt{\left(e^{\frac{-h}
{2}}\right)^{2}\emph{r}^{4}+8\emph{r}^{2}\emph{c}_{2}^{2}\emph{c}_{4}+16\emph{c}_{2}
^{2}}}{2\emph{c}_{2}\left(\emph{r}^{2}\emph{c}_{4}+2\right)}\big)}
\\\nonumber
&+&12\emph{c}_{2}\emph{c}_{3}\omega^2e^{\frac{-\lambda}{2}-\ln
\Big(\frac{-\emph{r}
^{2}e^{\frac{-h}{2}}+\sqrt{\left(e^{\frac{-h}{2}}\right)^{2}\emph{r}^{4}+8\emph{r}
^{2}\emph{c}_{2}^{2}\emph{c}_{4}+16\emph{c}_{2}^{2}}}{2\emph{c}_{2}\left(\emph{r}^{2}\emph{c}_{4}
+2\right)}\big)}\\\nonumber &+&\emph{c}_{2}^2
\omega^2\bigg\}^{\frac{1}{2}}\bigg\}\left\{2\emph{c}_{2}\emph{c}_{3}\left(3\omega^2
+1\right)\right\}^{-1}.
\end{eqnarray}
Figure \textbf{5} shows that upper face of the shape function
remains positive but the structure of WH is not asymptotically flat.
In the lower plane, WH throat is identified at $r_{0}=0.1$ and the
associated right plot leads to $\frac{db\left(r_{0}\right)}{dr}<1$.
Using Eq.(\ref{60}) in Eq.(\ref{21}), we have
\begin{equation}\nonumber
\mathrm{\rho}_{eff}+\mathrm{p}_{eff}-\textit{A} =
\frac{\emph{r}b'-b}{\emph{r}^3}.
\end{equation}
Figure \textbf{6} indicates that $\rho_{m}-\textit{A}\geq0$ and
$\mathrm{\rho}_{m}+\mathrm{p}_{m}-\textit{A}\geq0$ for $0\leq
\omega\leq1$ while $\mathrm{\rho}_{eff}+\mathrm{p}_{eff}-\textit{A}
\leq0$ for $-1\leq\omega\leq1$ which implies that viable traversable
wormhole solution exists in this particular range of $\omega$.
\begin{figure}
\epsfig{file=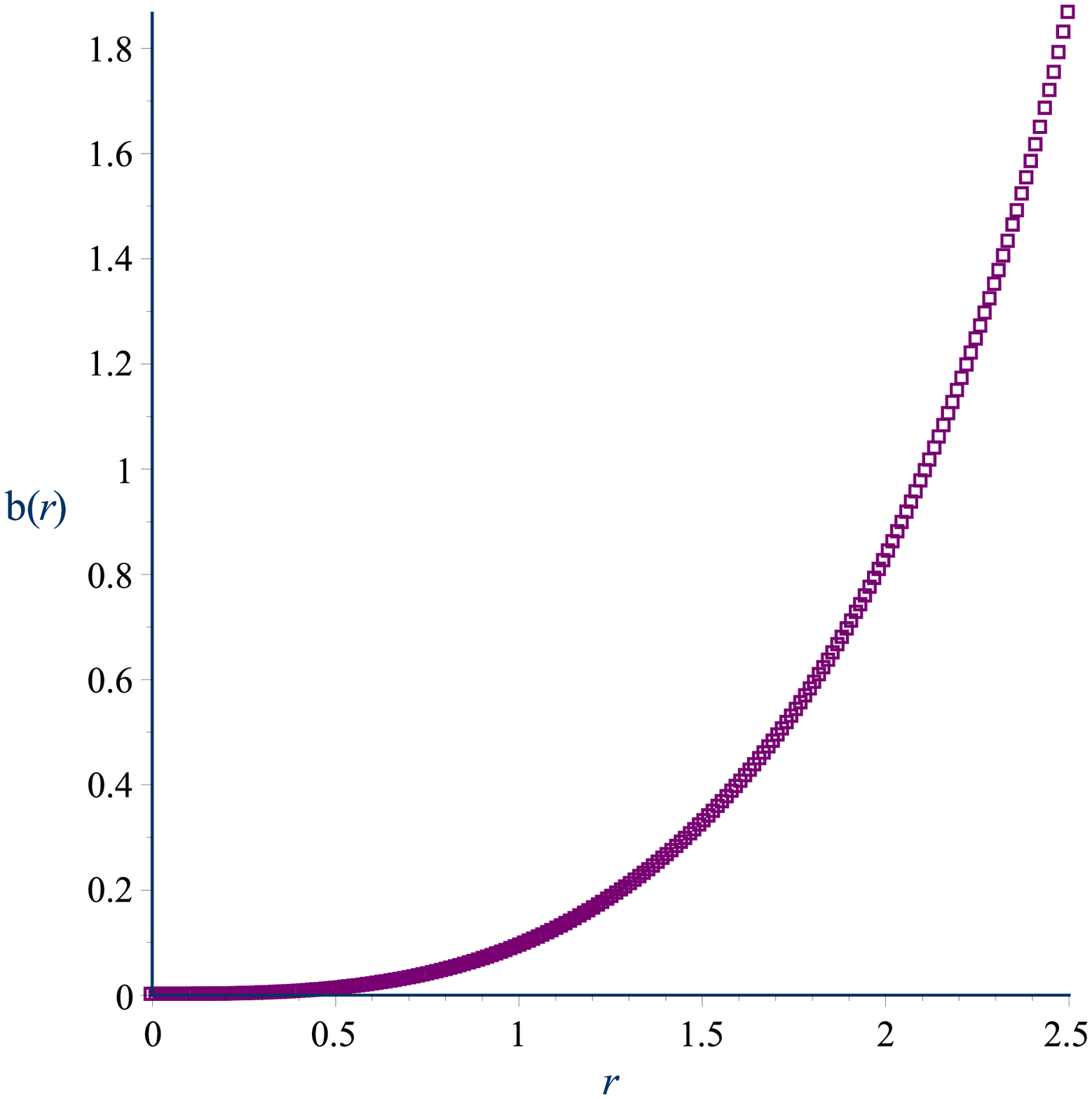,width=.5\linewidth}
\epsfig{file=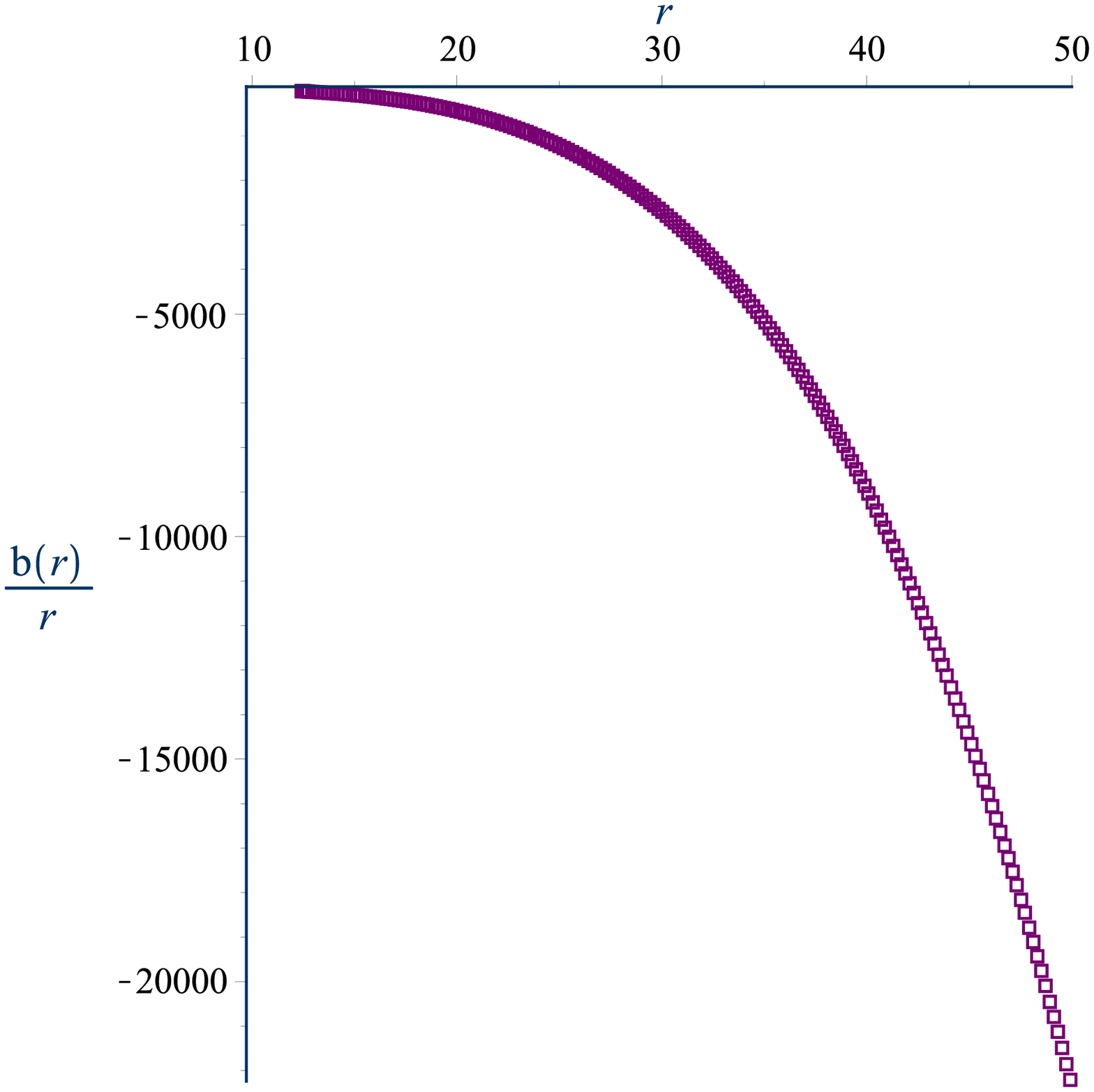,width=.5\linewidth}
\epsfig{file=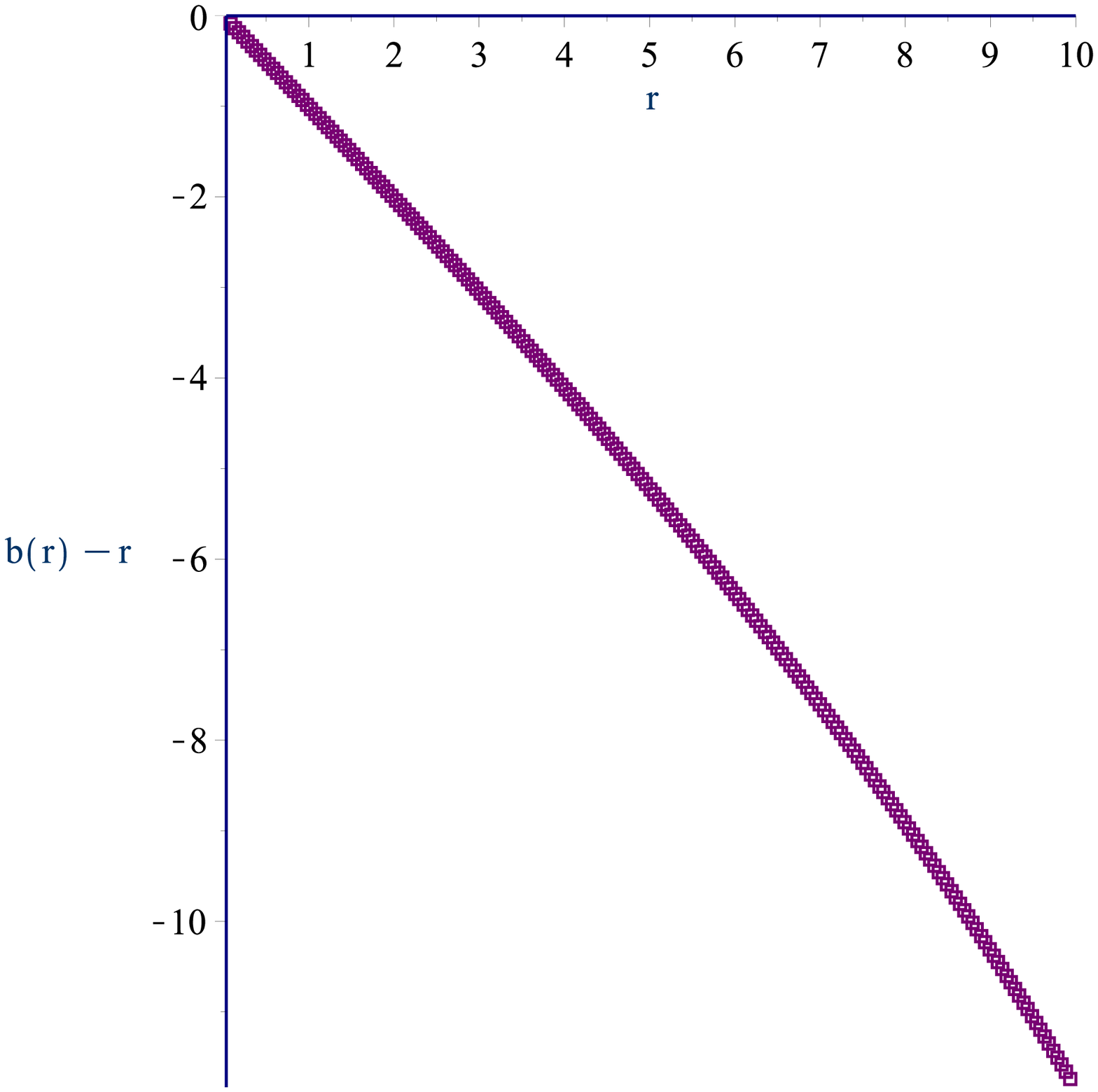,width=.5\linewidth}
\epsfig{file=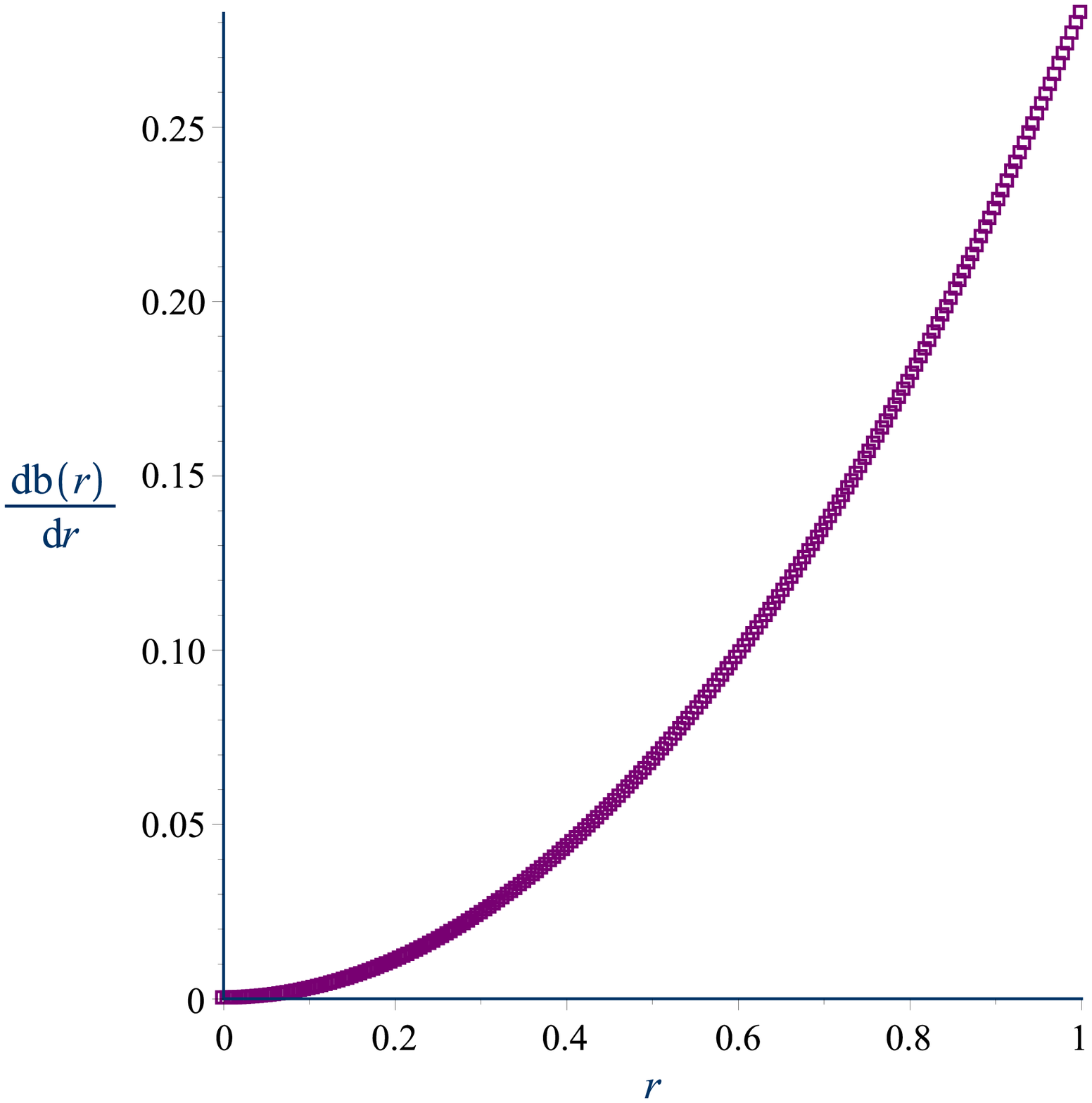,width=.5\linewidth} \caption{Graphs of
$b(\emph{r})$, $\frac{b(\emph{r})}{\emph{r}}$,
$b(\emph{r})-\emph{r}$ and $\frac{db(\emph{r})}{\emph{r}}$
corresponding to $\emph{r}$ for $\emph{c}_{2}$=5,
$\emph{c}_{4}$=-0.3 and h=1.}
\end{figure}
\begin{figure}
\epsfig{file=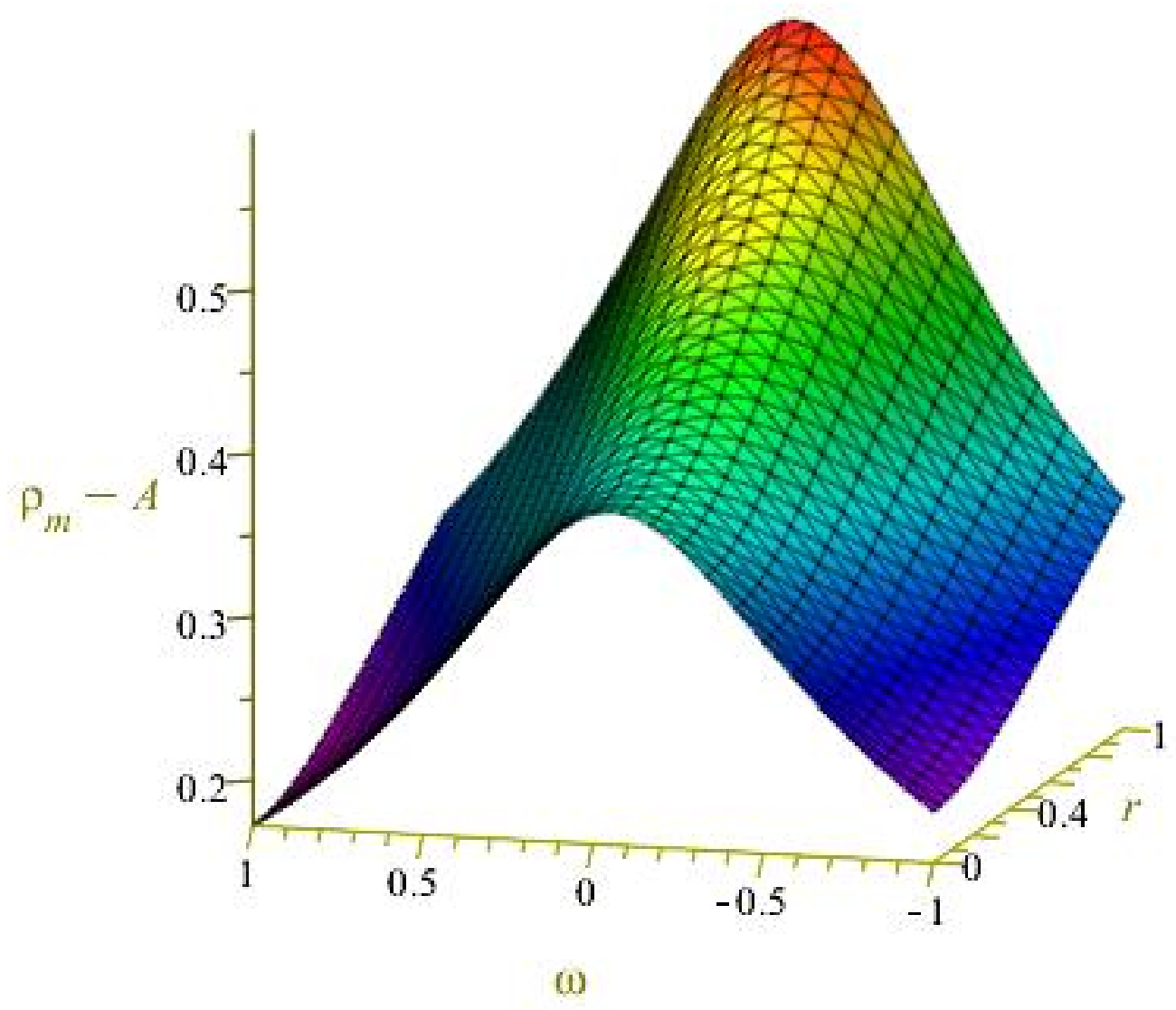,width=.5\linewidth}
\epsfig{file=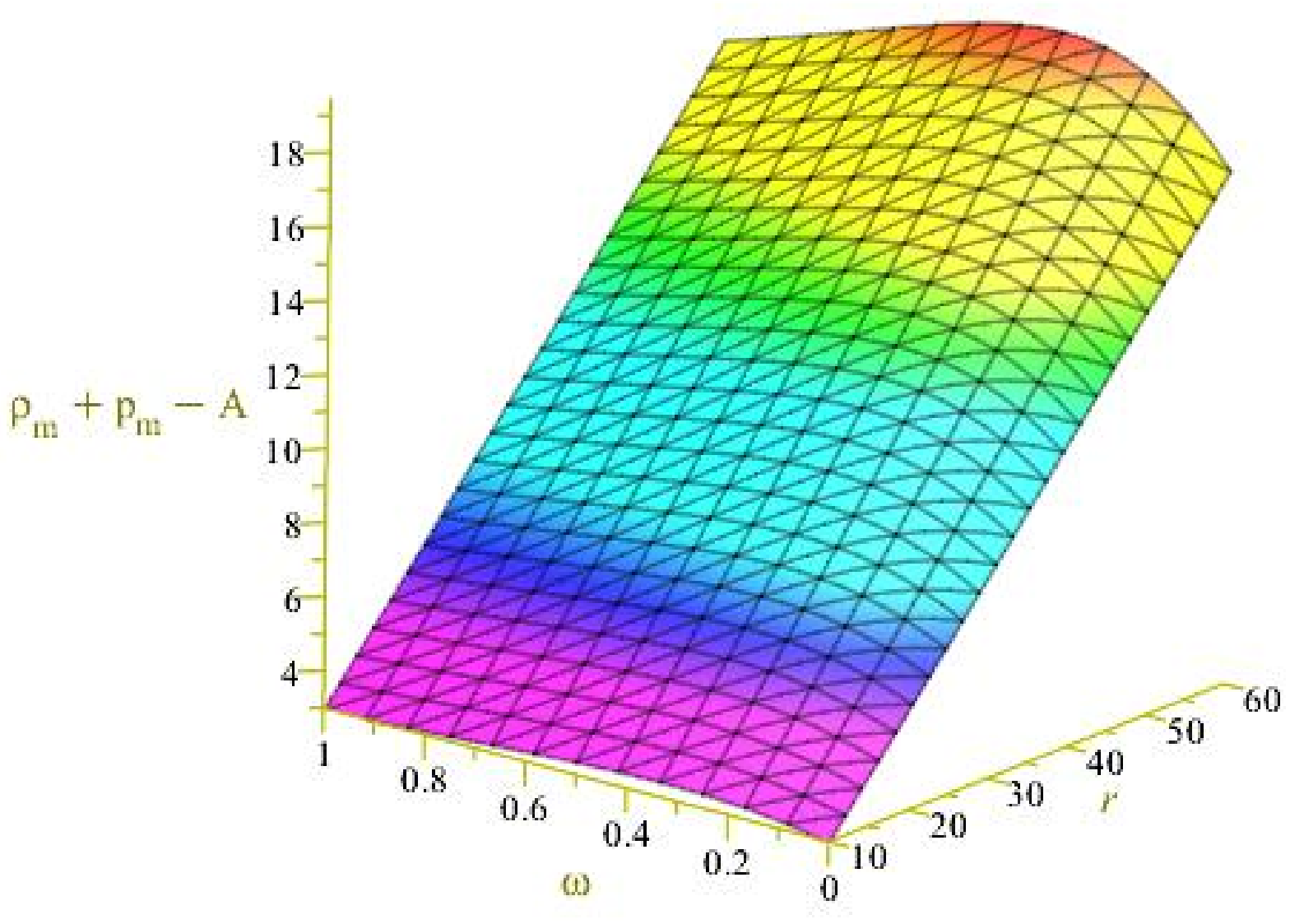,width=.5\linewidth} \center
\epsfig{file=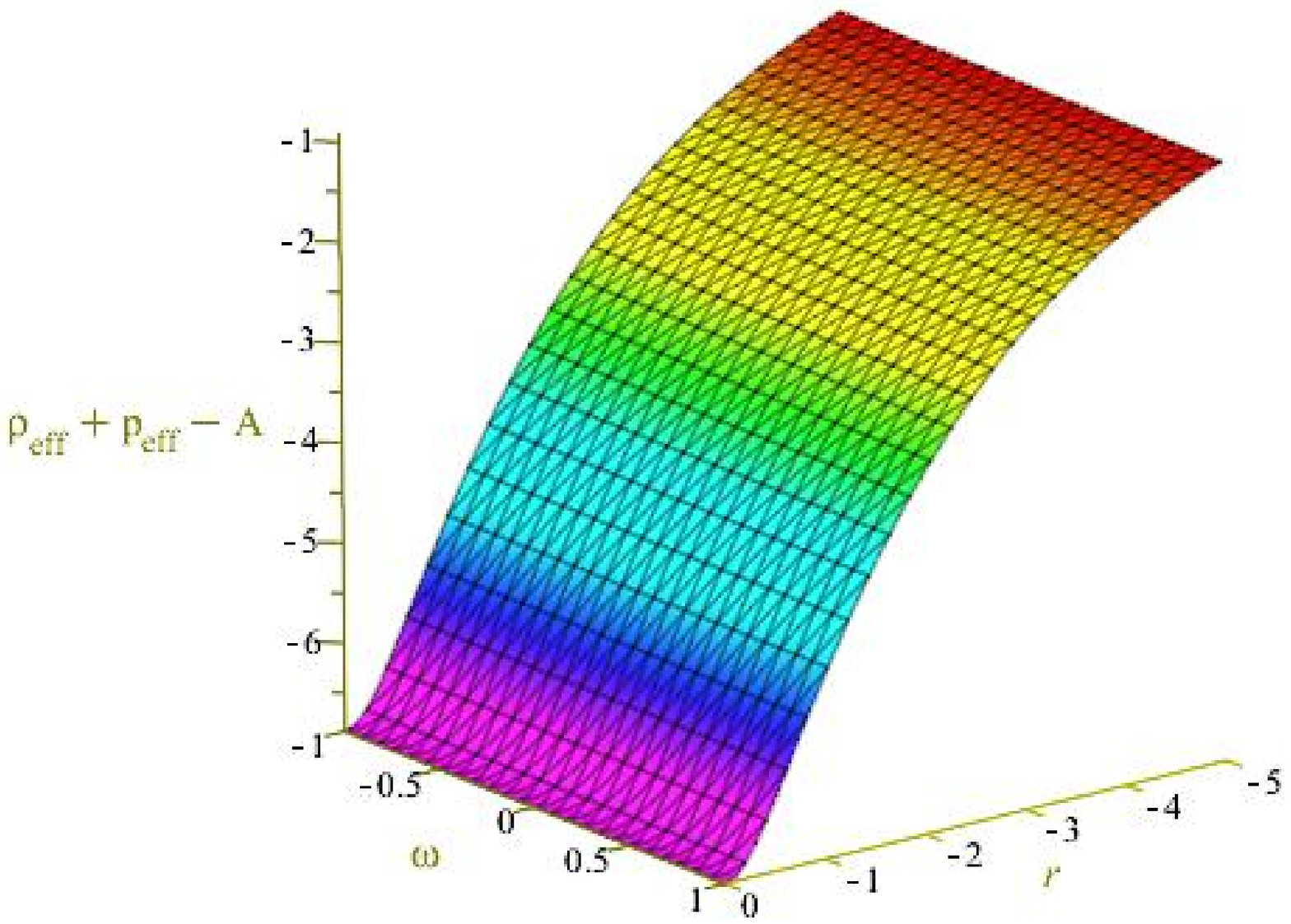,width=.5\linewidth} \caption{Graps of
$\mathrm{\rho}_{m}-\textit{A}$, $\mathrm{\rho}_{m}
+\mathrm{p}_{m}-\textit{A}$ and
$\mathrm{\rho}_{eff}+\mathrm{p}_{eff}-\textit{A}$ corresponding to
$\emph{r}$.}
\end{figure}

\subsubsection*{Case II: $\lambda(r)=-\frac{h}{\emph{r}}$}

Here, Eq.(\ref{59}) leads to
\begin{eqnarray}\nonumber
\vartheta(\emph{r}) &=& 2\ln
\bigg\{\left\{-\emph{r}^{3}e^{\frac{h}{2\emph{r}}}
+\left(\emph{r}^{6}e^{\frac{h}{\emph{r}}}+8h\emph{r}^{3}
\emph{c}_{2}^{2}\emph{c}_{4}+8r^{4}\emph{c}_{2}^{2}\emph{c}_{4}
\right.\right.\\\label{61}
&+&\left.\left.16h\emph{r}\emph{c}_{2}^{2}+16\emph{r}^2\emph
{c}_{2}^{2}\right)^{\frac{1}{2}}\right\}\left\{2\emph{c}_{2}
\left(\emph{r}^{2}\emph{c}_{4}+2\right)\right\}^{-1}\bigg\}.
\end{eqnarray}
The associated shape function is
\begin{eqnarray}\nonumber
b(\emph{r}) &=&
\bigg\{\left(\emph{r}^{5}e^{\frac{h}{\emph{r}}}+4h\emph{r}^{2}
\emph{c}_{2}^{2}\emph{c}_{4}+8h\emph{c}_{2}^{2}-2\emph{r}^{5}\emph{c}_{2}^{2}
\emph{c}_{4}^{2}-4\emph{r}^{3}\emph{c}_{2}^{2}\emph{c}_{4}
\right.\\\nonumber&-&\left.
\sqrt{\emph{r}^{6}e^{\frac{h}{\emph{r}}}+8h\emph{r}^{3}\emph{c}_{2}^{2}
\emph{c}_{4}+8\emph{r}^{4}\emph{c}_{2}^{2}\emph{c}_{4}-16h\emph{r}\emph{c}_{2}^{2}
+16\emph{r}^2\emph{c}_{2}^{2}} \right.\\\nonumber&\times&\left.
\emph{r}^{2}e^{\frac{h}{2\emph{r}}}\right)2\emph{r}^2\bigg\}\setminus
\bigg\{\left(8h\emph{r}^{3}\emph{c}_{2}^{2}\emph{c}_{4}+8\emph{r}^{4}\emph
{c}_{2}^{2}\emph{c}_{4}-16h\emph{r}\emph{c}_{2}^{2}+\emph{r}^2
\right.\\\nonumber&\times&\left.
16\emph{c}_{2}^{2}+\emph{r}^{6}e^{\frac{h}{\emph{r}}}\right)^{\frac{1}{2}}
-\emph{r}^{3}e^{\frac{h}{2\emph{r}}}\bigg\}^{2}.
\end{eqnarray}
Figure \textbf{7} shows that $b(r)$ remains positive but the
geometry of WH is not asymptotically flat and WH throat is located
at $\emph{r}_{0} = 2$ with $\frac
{db\left(\emph{r}_{0}\right)}{d\emph{r}}<1$. Figure \textbf{8}
exhibits that $\rho_{m} -\textit{A}\geq0$ and
$\mathrm{\rho}_{m}+\mathrm{p}_{m}-\textit{A}\geq0$ for
$-1\leq\omega\leq1$ whereas
$\mathrm{\rho}_{eff}+\mathrm{p}_{eff}-\textit{A}<0$ for
$-1\leq\omega\leq0$, implying that physically viable and traversable
WH exists.
\begin{figure}
\epsfig{file=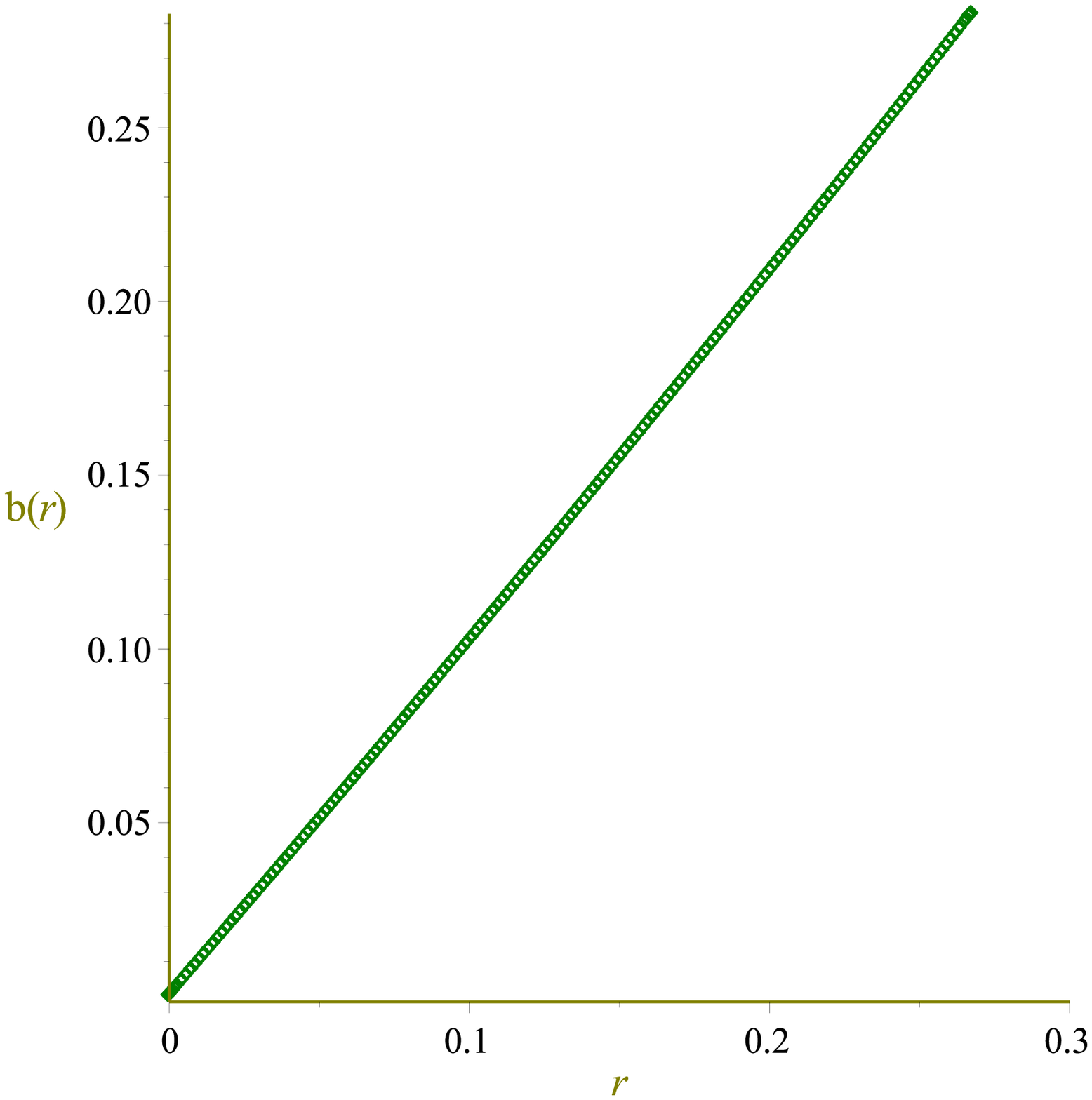,width=.5\linewidth}
\epsfig{file=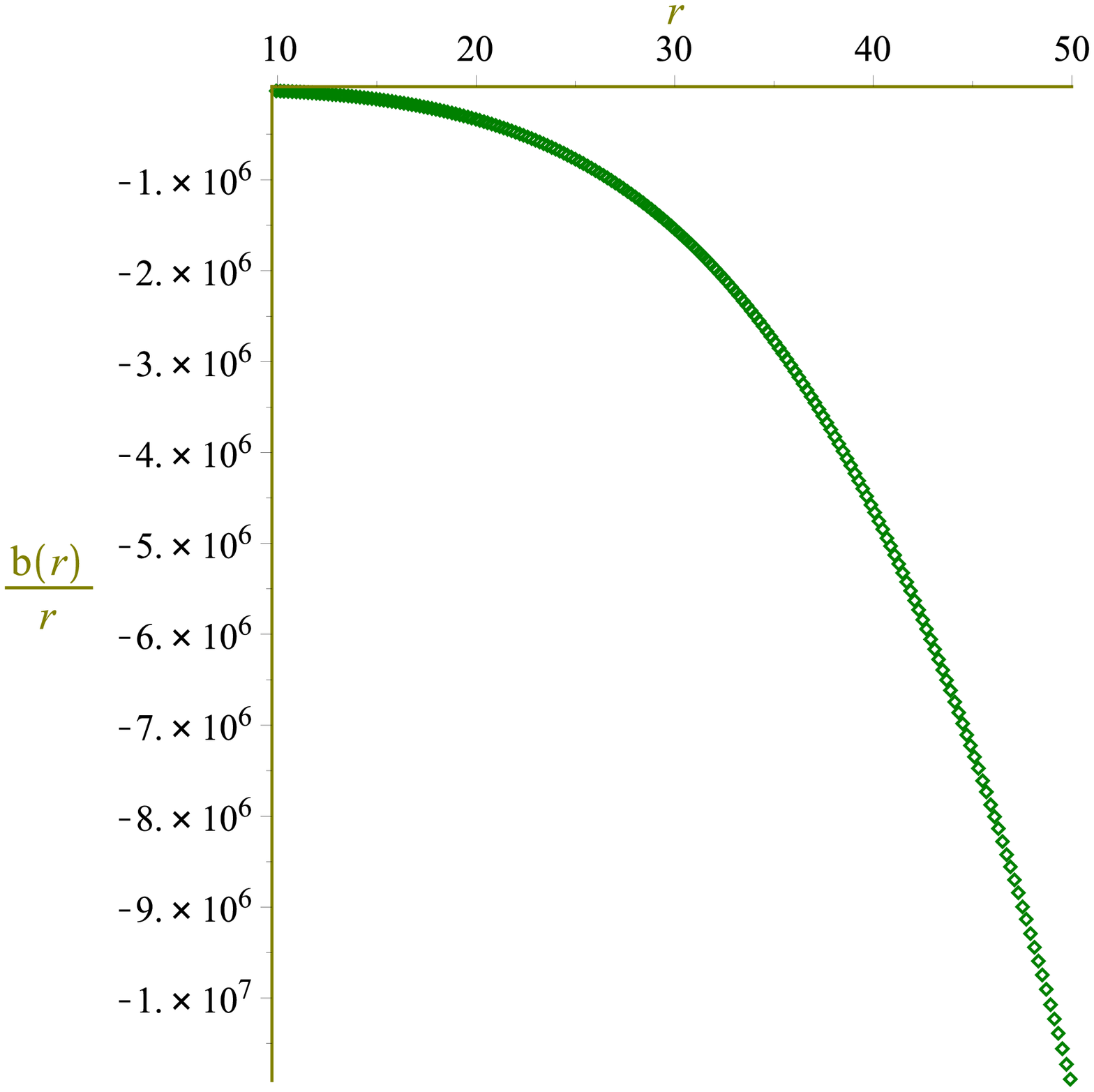,width=.5\linewidth}
\epsfig{file=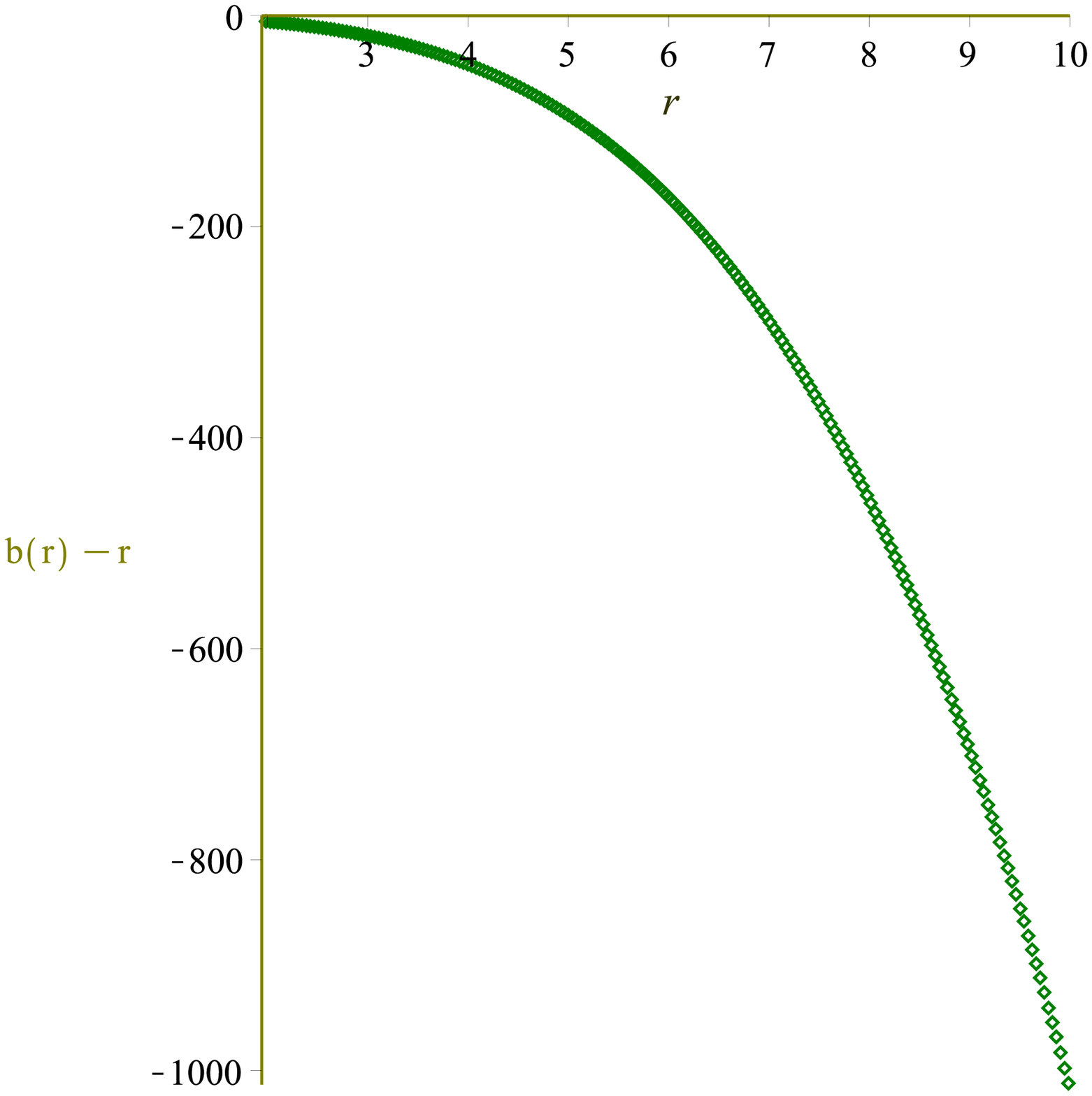,width=.5\linewidth}
\epsfig{file=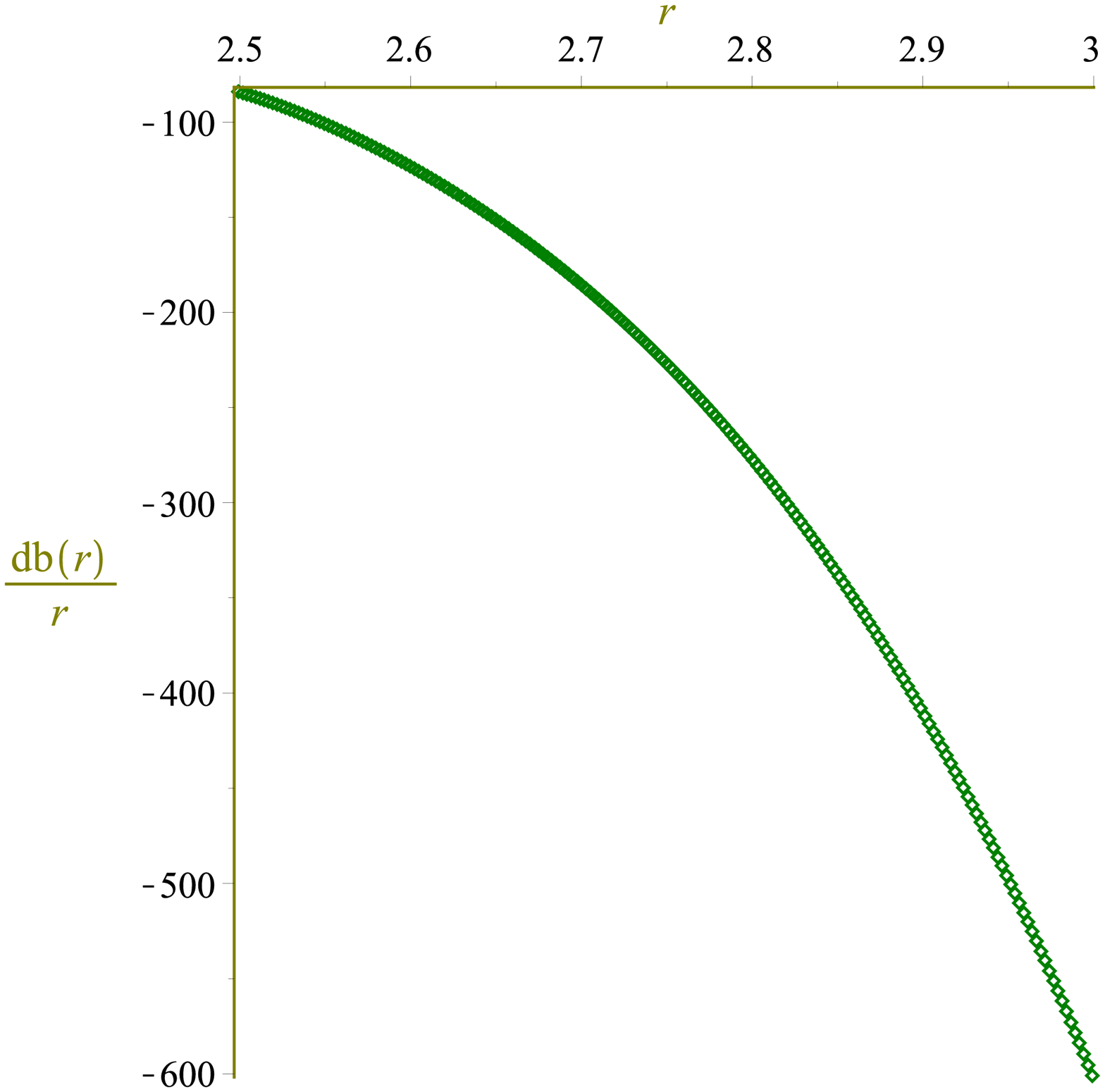,width=.5\linewidth} \caption{Graphs of
$b(\emph{r})$, $\frac{b(\emph{r})}{\emph{r}}$,
$b(\emph{r})-\emph{r}$ and $\frac{db(\emph{r})}{\emph{r}}$
corresponding to $\emph{r}$ for $\emph{c}_{2}$=0.4,
$\emph{c}_{4}$=0.2 and h=-5.}
\end{figure}
\begin{figure}
\epsfig{file=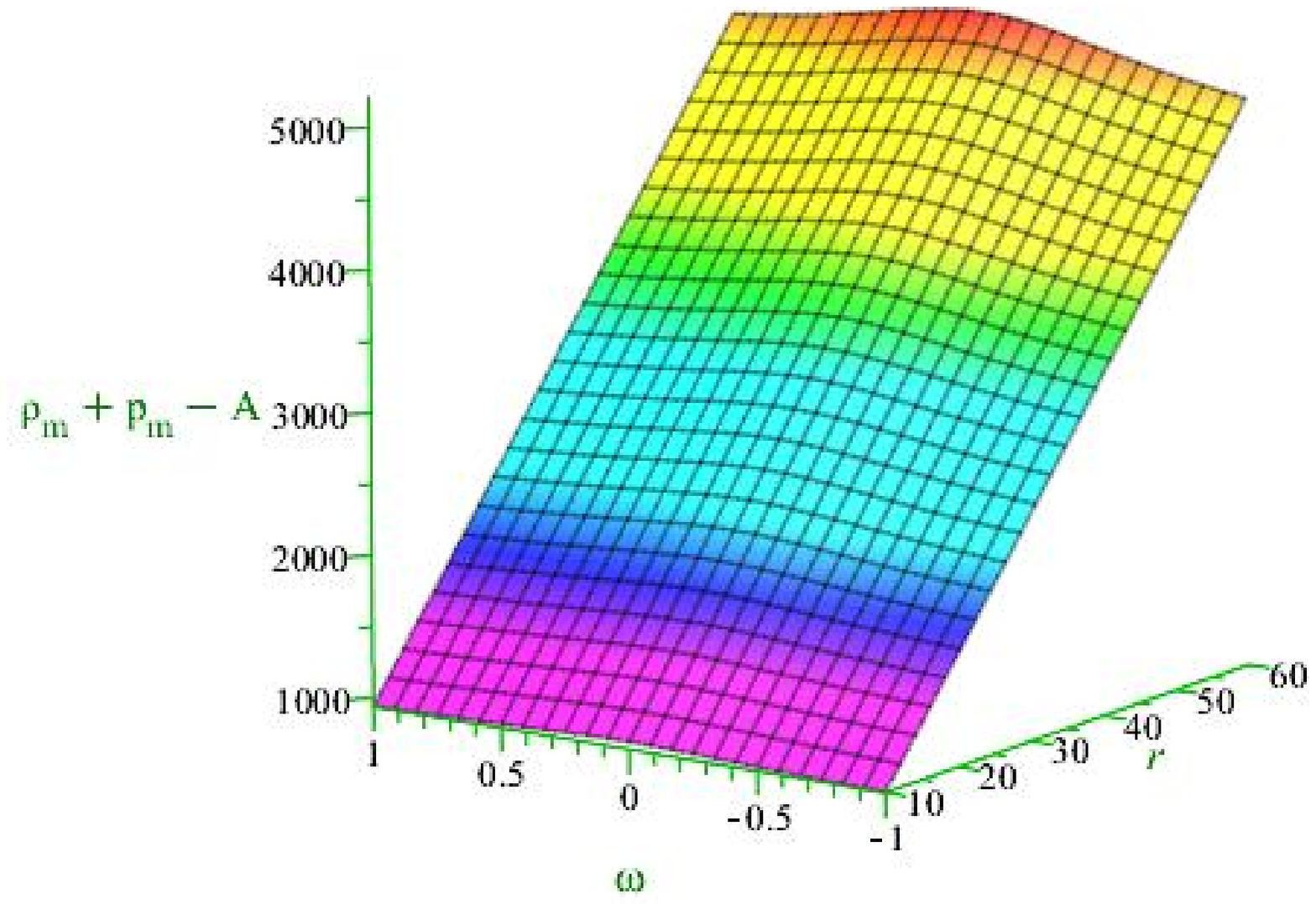,width=.5\linewidth}
\epsfig{file=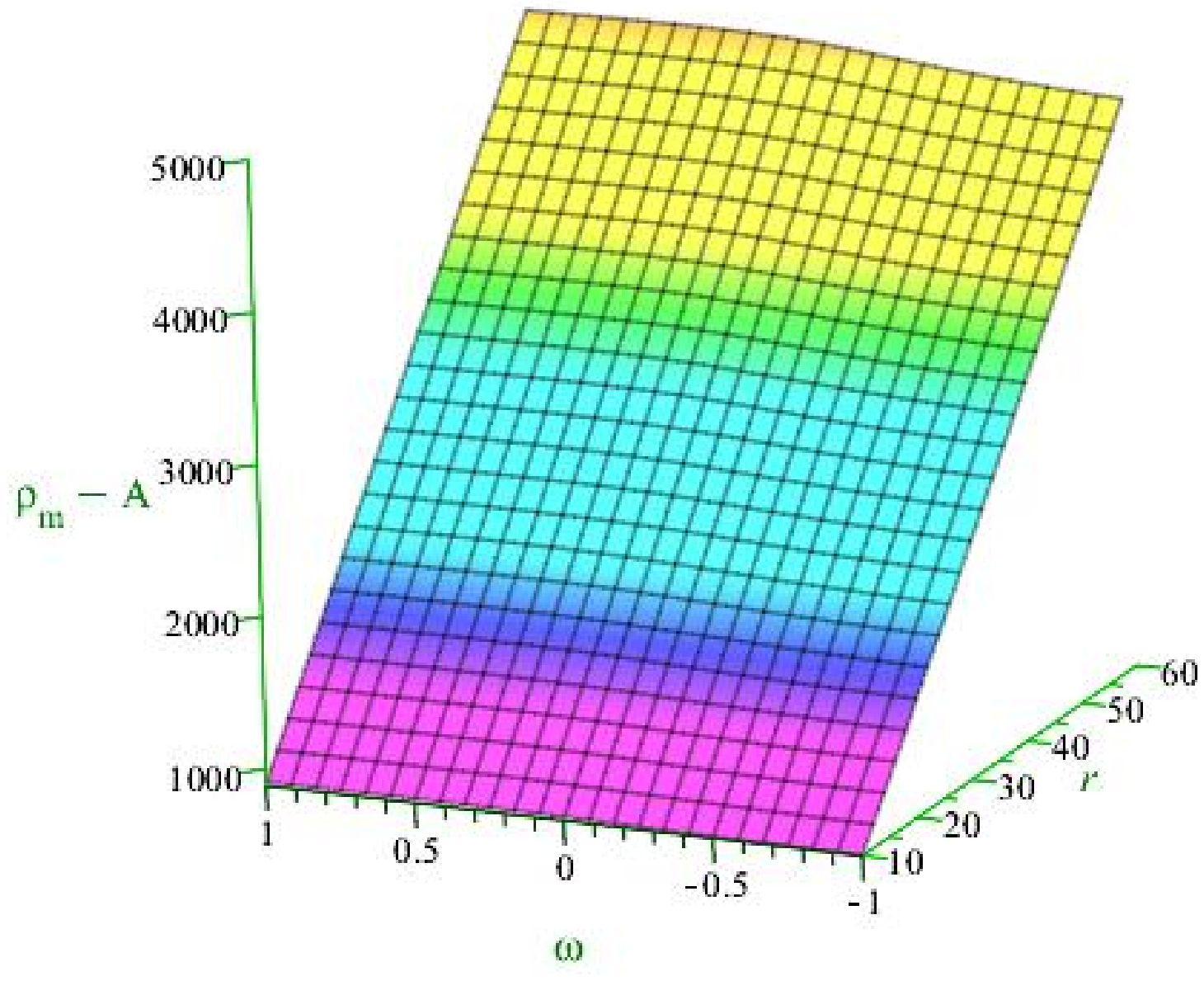,width=.5\linewidth} \center
\epsfig{file=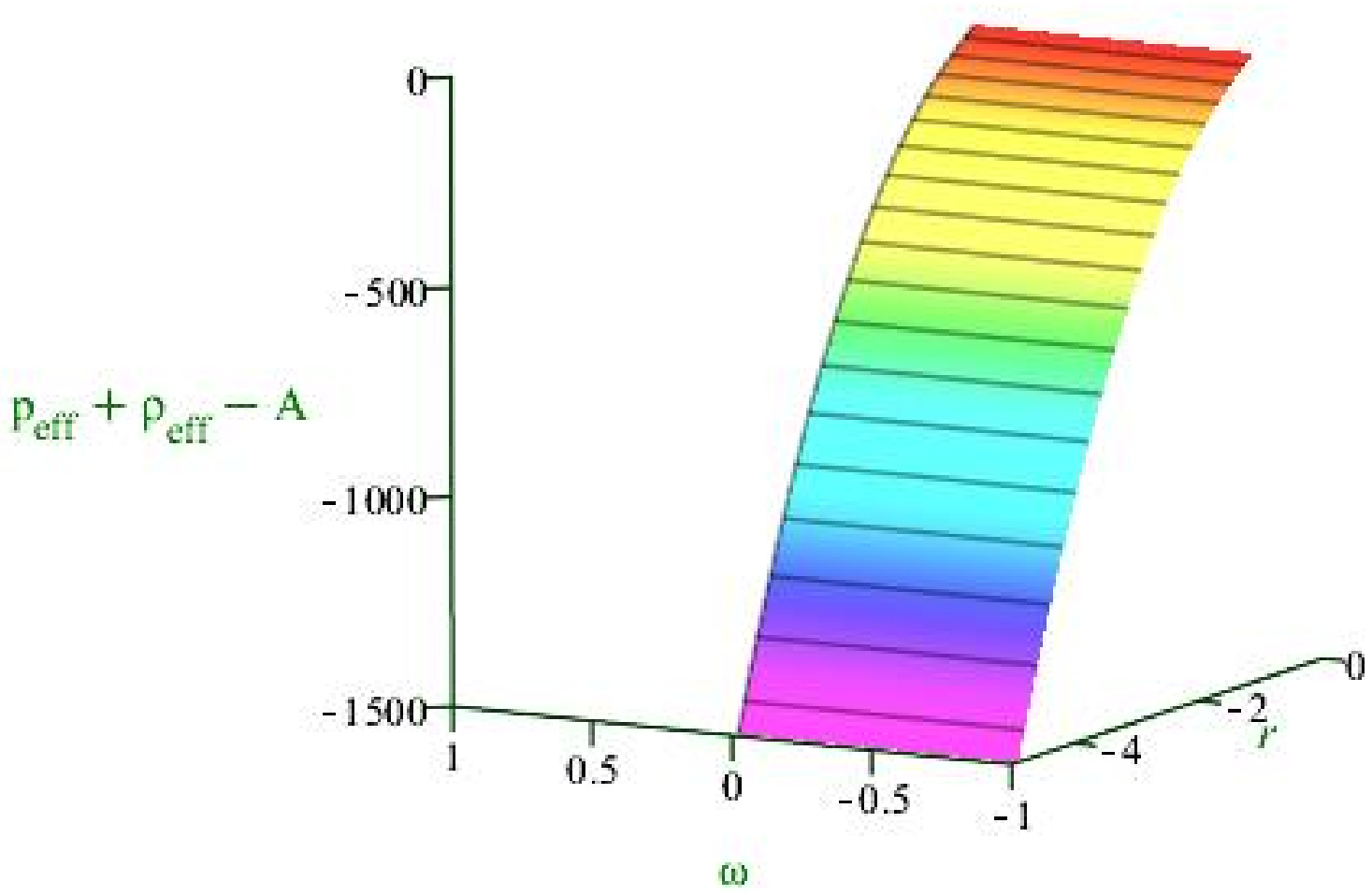,width=.5\linewidth} \caption{Graphs of
$\mathrm{\rho}_{m}-\textit{A}$, $\mathrm{\rho}_{m}
+\mathrm{p}_{m}-\textit{A}$ and
$\mathrm{\rho}_{eff}+\mathrm{p}_{eff} -\textit{A}$ corresponding to
$\emph{r}$.}
\end{figure}

\section{Stability Analysis}
\begin{figure}
\epsfig{file=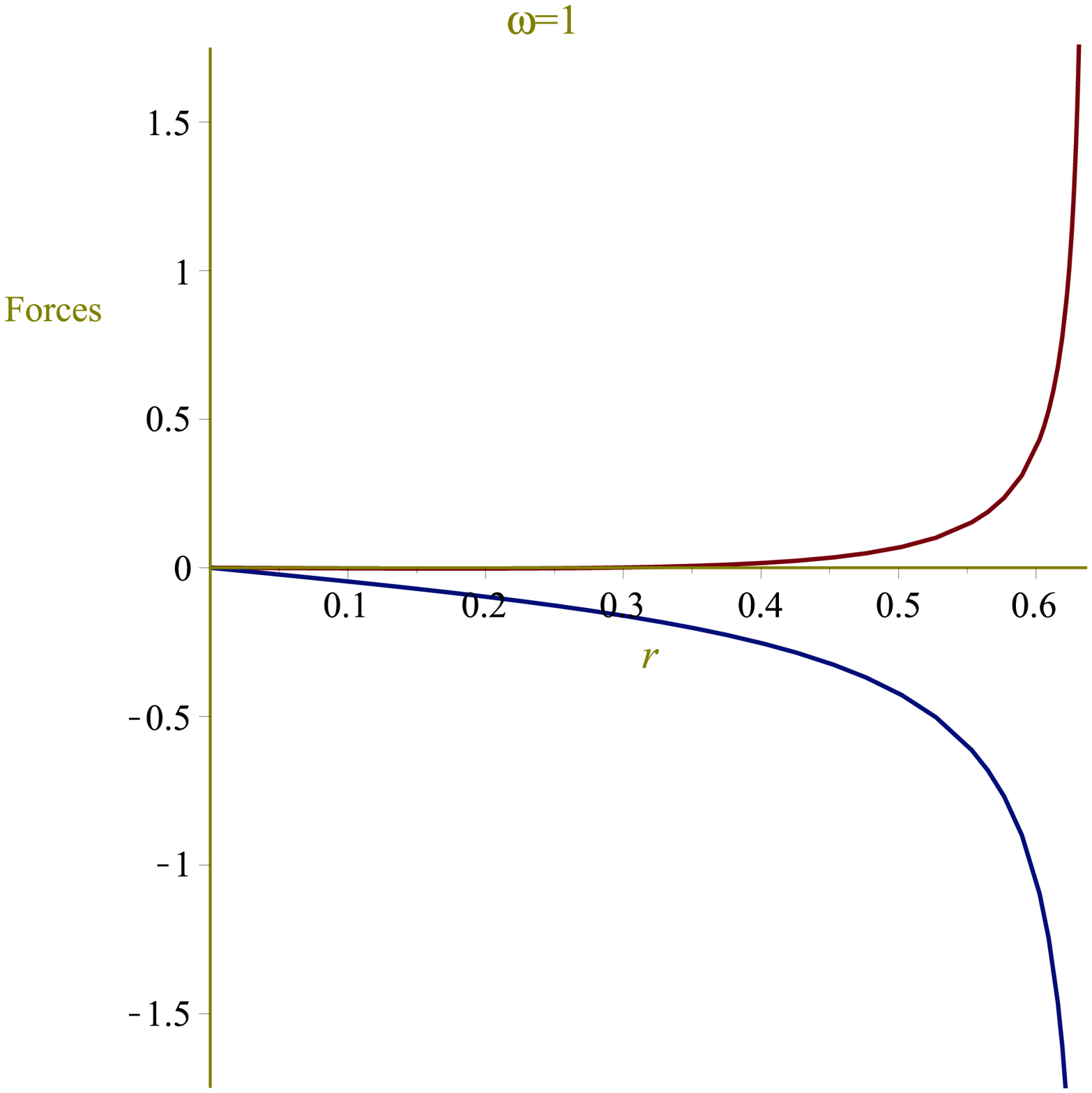,width=.5\linewidth}
\epsfig{file=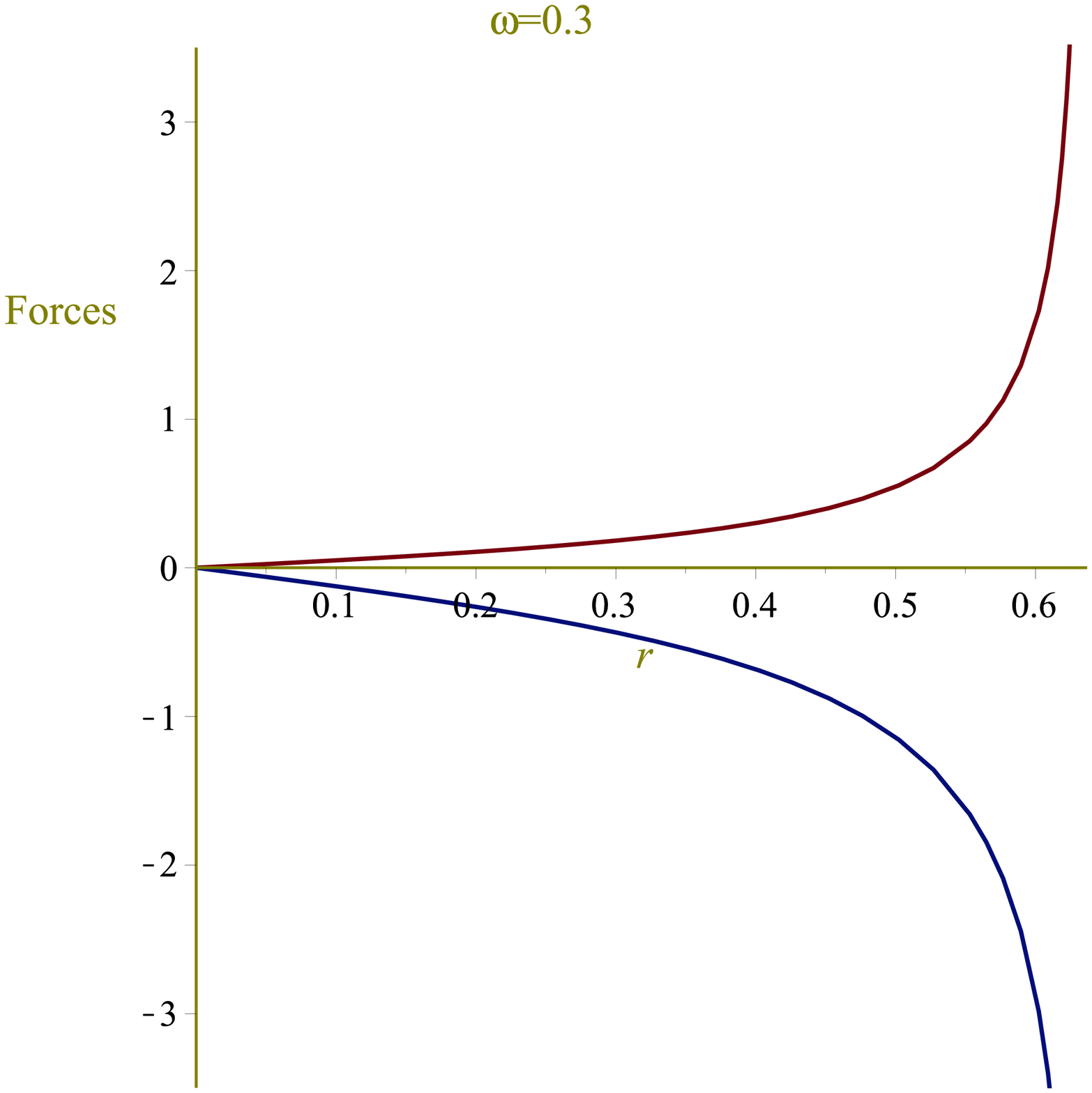,width=.5\linewidth}
\epsfig{file=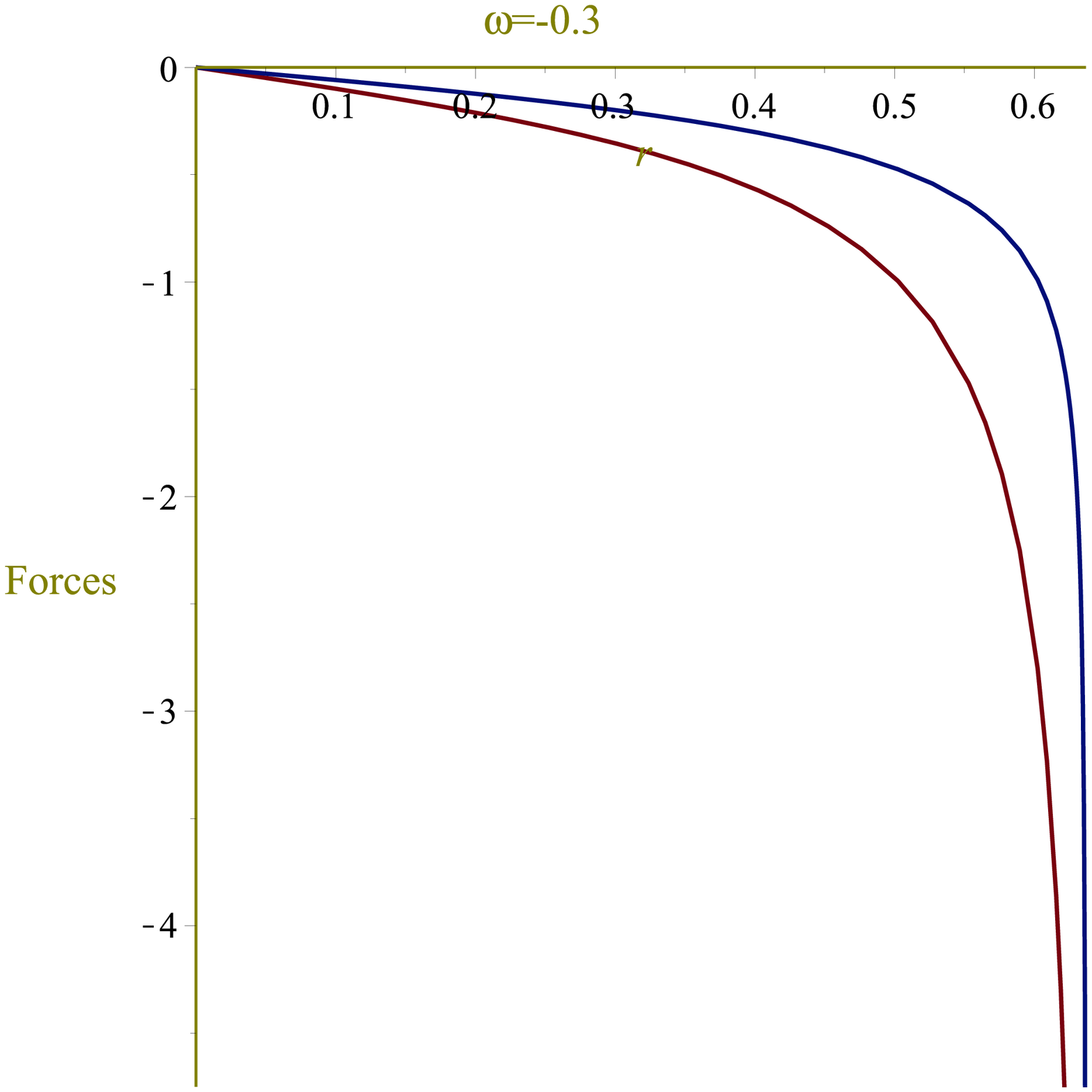,width=.5\linewidth}
\epsfig{file=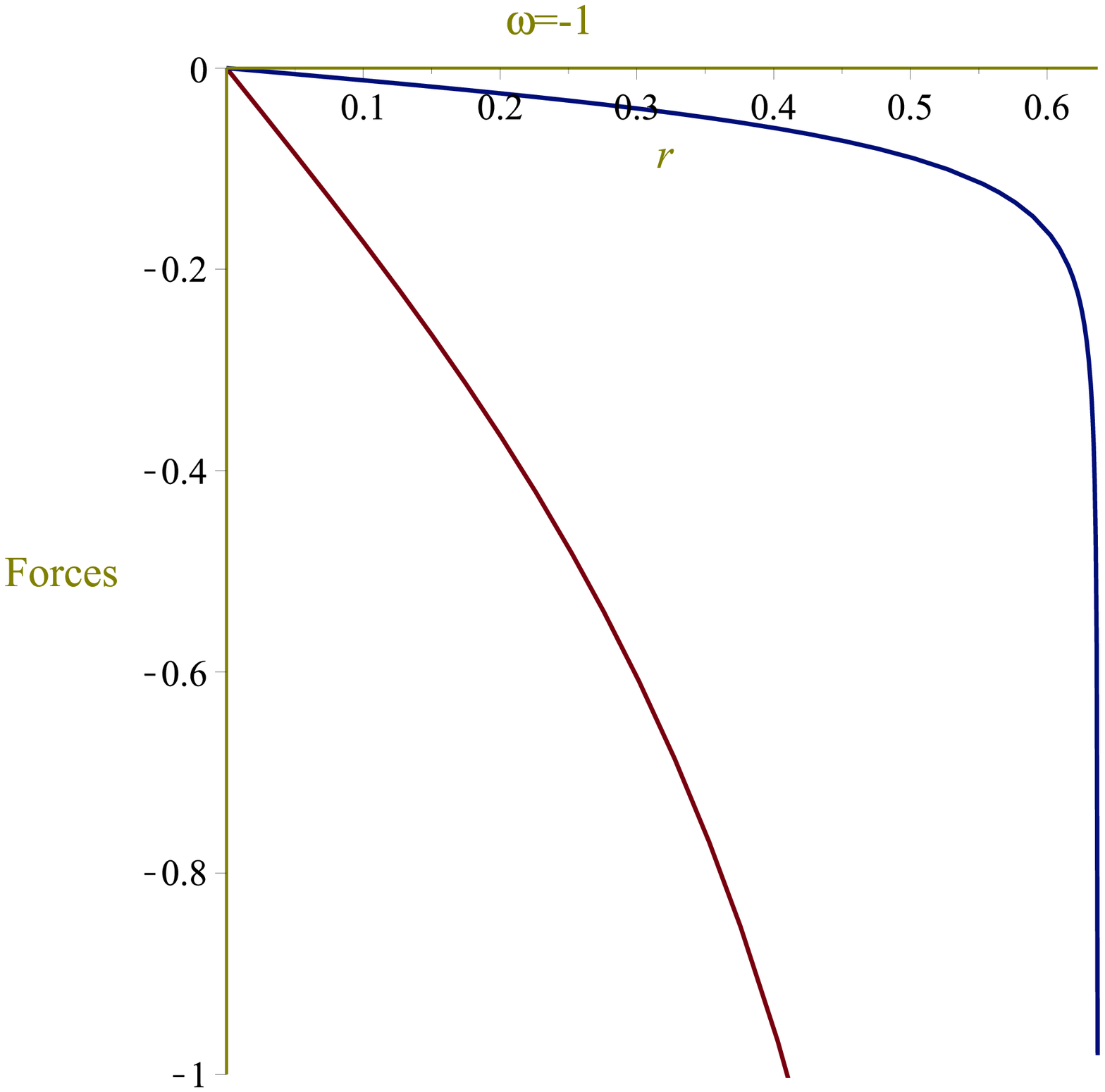,width=.5\linewidth} \caption{Graphs of
$\mathcal{F}_{g}$ (red) and $\mathcal{F}_{\mathfrak{h}}$ (blue)
versus $\emph{r}$ for constant red-shift function with
$\emph{c}_{2}$=-0.5, $\emph{c}_{3}$=0.1, $\emph{c}_{4}$=-5 and h=1.}
\end{figure}
\begin{figure}
\epsfig{file=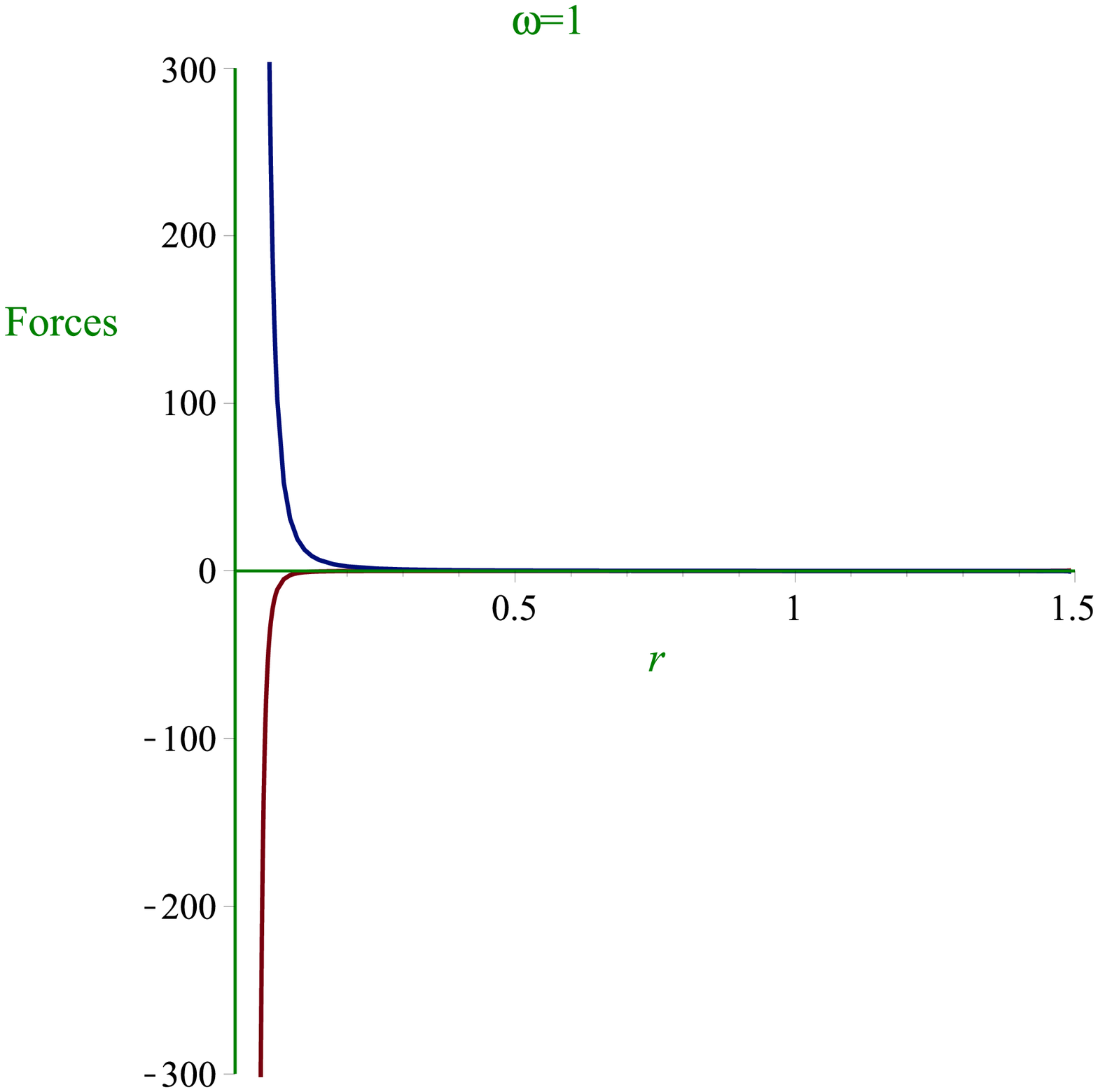,width=.5\linewidth}
\epsfig{file=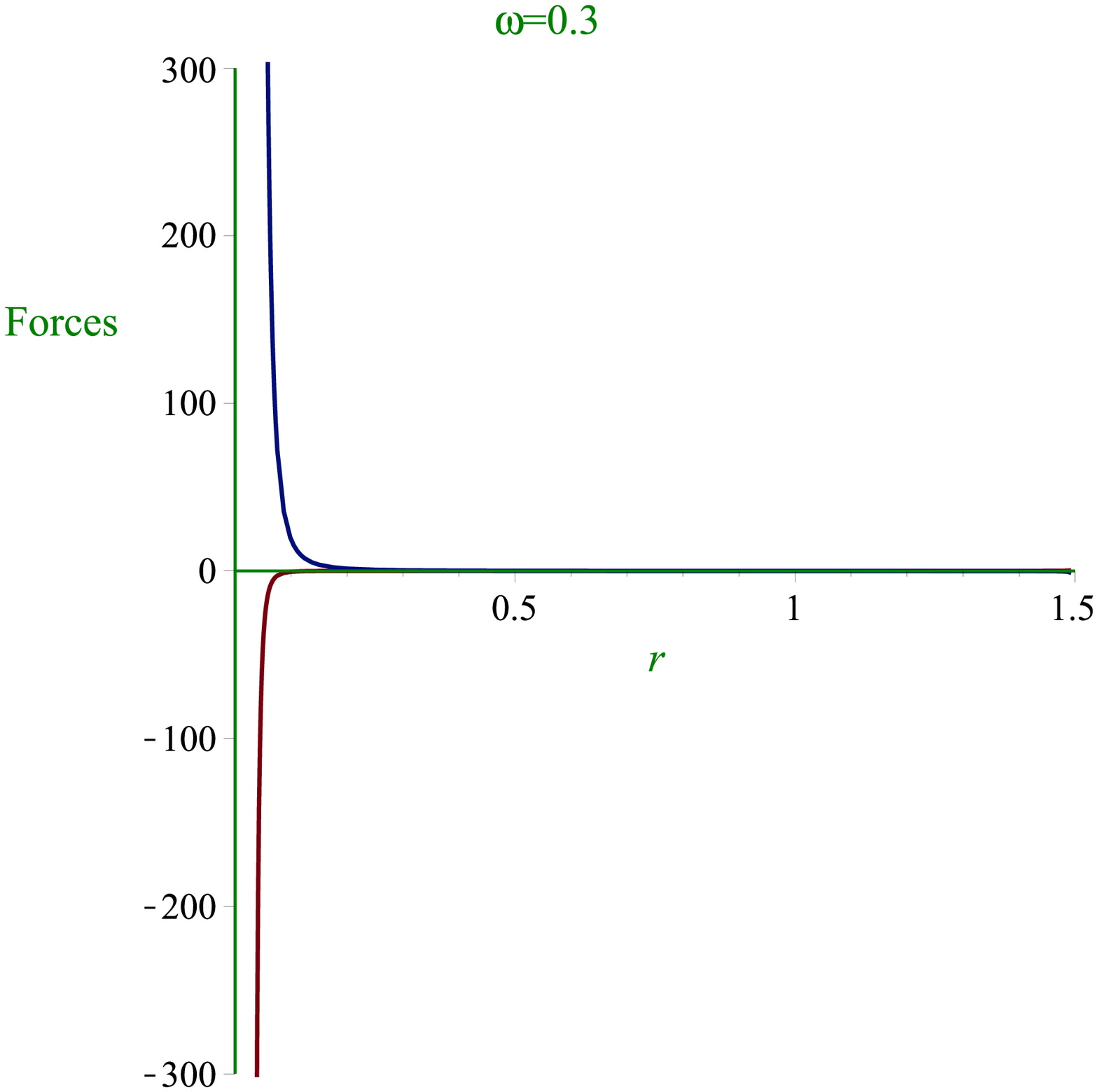,width=.5\linewidth}
\epsfig{file=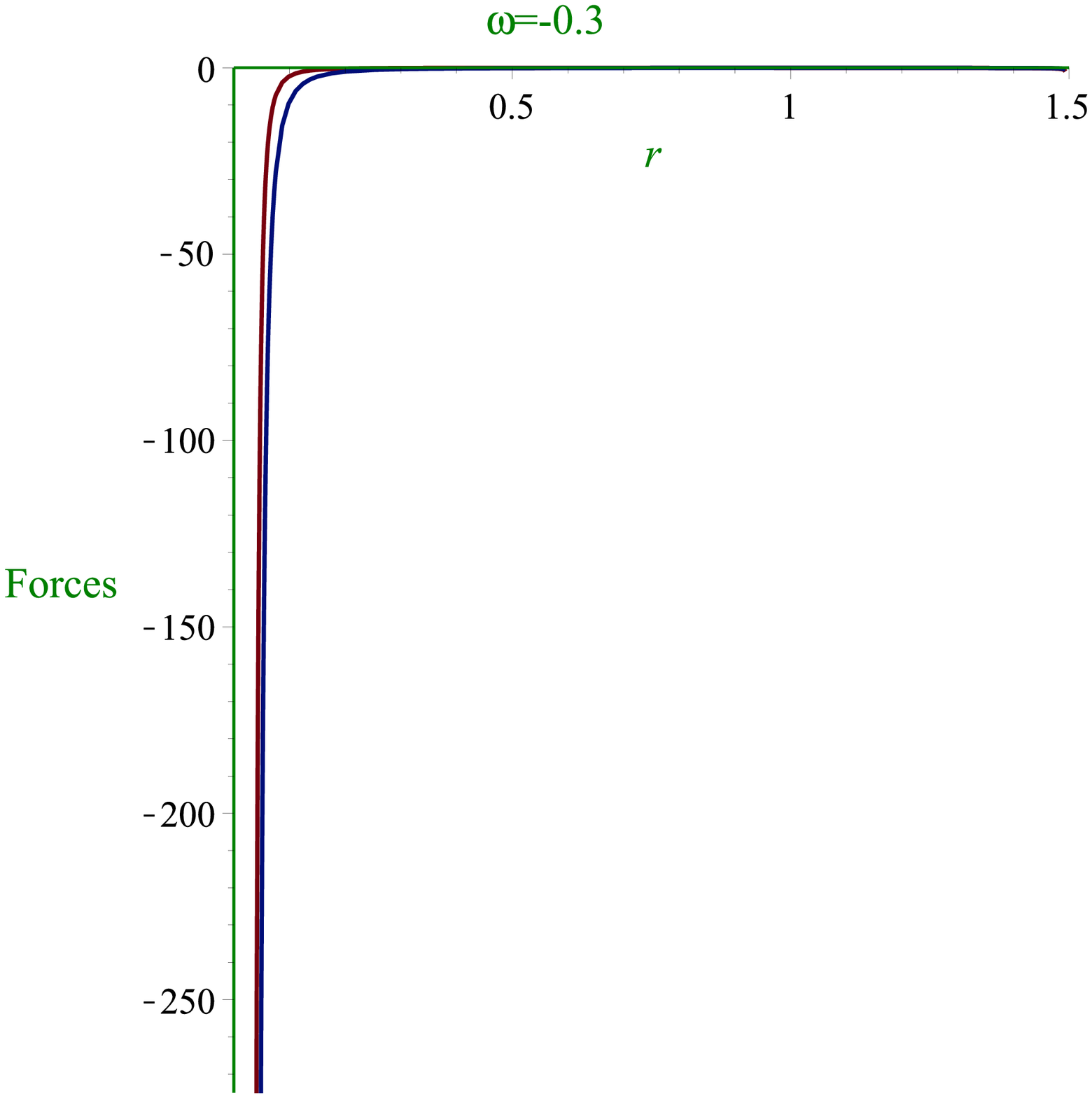,width=.5\linewidth}
\epsfig{file=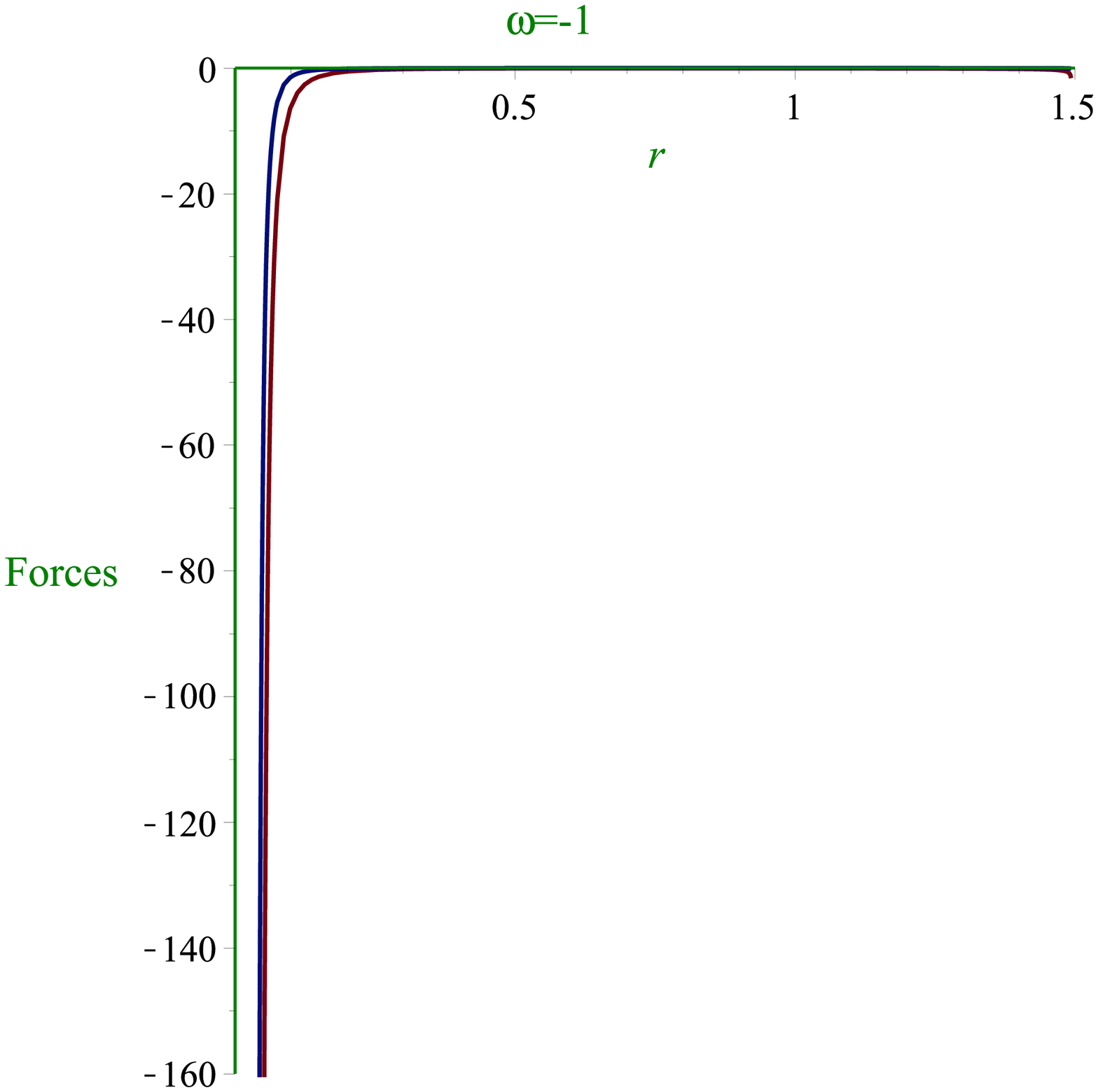,width=.5\linewidth} \caption{Graphs of
$\mathcal{F}_{g}$ (red) and $\mathcal{F}_{\mathfrak{h}}$ (blue)
versus $\emph{r}$ for variable red-shift function with
$\emph{c}_{2}$=-8, $\emph{c}_{3}$=0.9, $\emph{c}_{4}$=-0.9 and
h=0.5.}
\end{figure}

Here, we investigate the stability of viable and traversable WH
solutions for both choices of red-shift function by using the TOV
equation. We take into account non-conserved stress-energy tensor
and formulate TOV equation for isotropic matter configuration as
\begin{equation}\label{62}
\mathrm{p}_{m}'+f_{\mathbf{T}^{2}}\left\{\left(3\mathrm{p}_{m}\mathrm{p}_{m}'
+\mathrm{\rho}_{m}\mathrm{\rho}_{m}'\right)+\frac{\lambda'}{2}\left(3\mathrm{p}
_{m}^2+\mathrm{\rho}_{m}^2+4\mathrm{p}_{m}\mathrm{\rho}_{m}\right)\right\}
+\frac{\lambda'}{2}\left(\mathrm{\rho}_{m}+\mathrm{p}_{m}\right)=0.
\end{equation}
This equation demonstrates the combination of gravitational force
$\left(\mathcal{F}_{g}\right)$ and hydrostatic force
$\left(\mathcal{F}_{h}\right)$ that determine the equilibrium state
of WH. In the light of Eq.(\ref{62}), these forces defined as
\begin{eqnarray}\label{63}
&&\mathcal{F}_{\mathfrak{h}}
=\mathrm{p}_{m}'\left(1+3\mathrm{p}_{m}f_{\mathbf{T}^{2}}\right),
\\\label{64}
&&\mathcal{F}_{g} =
\frac{\lambda'}{2}\left\{\left(\mathrm{\rho}_{m}+\mathrm{p}_{m}
\right)+f_{\mathbf{T}^{2}}\left(3\mathrm{p}_{m}^2+\mathrm{\rho}
_{m}^2+4\mathrm{p}_{m}\mathrm{\rho}_{m}\right)+f_{\mathbf{T}^{2}}
\mathrm{\rho}_{m}\mathrm{\rho}_{m}'\right\}.
\end{eqnarray}
The null impact of these forces
$\left(\mathcal{F}_{\mathfrak{h}}+\mathcal{F}_{g}= 0\right)$ ensure
the presence of stable traversable WH. Figure \textbf{9} shows the
stable and unstable behavior of viable traversable WH with constant
red-shift function at distinct evolutionary eras. In the upper face,
both plots indicate the stable state of WH for $\omega=1$ and
$\omega=0.3$. This exhibits that WH preserves its stable state in
the stiff matter era that remains until the radiation dominated era.
The paths of both forces in the lower face are identical in
direction as well as magnitude and hence violates the condition of
equilibrium for both $\omega=-1$ and $\omega=-0.3$. This leads to
the presence of stable and realistic traversable WH in the
decelerating era whereas this stable state is disturbed in the
cosmic accelerated expansion phase. For
$\lambda(\emph{r})=\frac{-h}{\emph{r}}$, Eqs.(\ref{54}) and
(\ref{63}) describe the stable state of WH incorporating with stiff
matter, radiation dominated era and dark energy phase. The upward
and downward faces of Figure \textbf{10} explain the fate of
traversable WH such that in the decelerating phase it admits stable
state whereas unstable state occurs in DE era.

\section{Concluding Remarks}

Noether symmetries are not just a mechanism to deal with the
dynamical solutions, but also their possible existence may provide
some feasible conditions so that one can choose some viable universe
models according to recent observations. Lagrangian multipliers are
useful to re-shape the Lagrangian into its canonical form which may
prove to be quite useful to reduce the dynamics of the system and
eventually help in determining the exact solutions. The existence of
Noether charges are considered important in the literature and
conserved quantities play an important role to analyze the
mysterious universe.

The main challenge whether a WH exists is usually based on energy
conditions which appears to be a fascinating subject in gravitation.
In GR, the fundamental constituent for the existence of physically
viable WH is the violation of energy conditions due to the presence
of exotic matter. Modified gravitational theories have received
significant attention as a possible alternative to GR during the
last few decades. Many researchers found this quite significant to
examine whether different modified theories violate the energy
conditions by the effective energy-momentum tensor which leads to
exotic matter and hence confirms the existence of a physically
viable WH.

In this paper, we have used the Noether symmetry technique to
evaluate some exact solutions that help to construct static WHs in
EMSG and also investigate whether ordinary matter assists WHs or not
in this theory. We have discussed the presence of exotic and normal
matter in WHs through effective and ordinary energy bounds. We have
taken a minimal coupling model to examine the viable WH geometry for
both dust as well as non-dust matter distribution. We have also
checked the stable and unstable states of these WH solutions through
the TOV equation. We have formulated the complicated system through
the Noether symmetry technique and determined the generators of
symmetry with corresponding conserved quantities in the presence of
shape function and energy density.

For EMSG model, we have examined the viability of WH solutions with
red-shift functions $\lambda(\emph{r})=h$ and
$\lambda(\emph{r})=-\frac{h}{\emph{r}}$ for dust as well as non-dust
matter distribution and evaluated exact solutions. It is found that
for $\lambda(\emph{r})=h$, WH fulfills all the necessary conditions
for dust fluid while for non-dust distribution, WH does not preserve
asymptotically flat behavior. The energy density for normal matter
in both cases remains positive whereas the effective stress-energy
tensor violates the $\mathbb{NEC}$. This implies that traversable WH
exists whereas the existence of normal matter gives physically
realistic WH. For $\lambda(\emph{r})=-\frac{h}{\emph{r}}$, all
necessary conditions of WHs are satisfied for both matter
distributions and specific relation between matter variables is
considered in non-dust case
($\mathrm{p}_{m}=\omega\mathrm{\rho}_{m}$). For both matter
distributions, we have found $\mathrm{\rho}_{m}-\textit{A}\geq0$,
$\mathrm{\rho}_{m}+\mathrm{p}_{m}-\textit{A} \geq0$ and
$\mathrm{\rho}_{eff}+\mathrm{p}_{eff}-\textit{A}\leq0$. These
inequalities indicate the presence of physically viable and
traversable WH. Finally, we have checked the stability of WH against
stiff matter-dominated and radiation-dominated era for both values
of the red-shift function. This stable state of WHs becomes unstable
as the universe passes through dust dominated phase and enters into
the dark energy era.

Lobo and Oliveira \cite{38} discussed the WH geometry in $f(R)$
gravity and found that no viable wormhole solution exists for the
vacuum case. Zubair et al. \cite{39} found static WH solutions with
anisotropic, isotropic, and barotropic matter contents in $f(R,T)$
gravity. For this purpose, they considered a generalization of
Starobinsky $f(R)$ model with linear form of $f(T)$ and tackled
complexity of the field equations via numerical approach. To analyze
the physical viability of WHs, they constructed a graphical analysis
of energy bounds for all considered fluids and found that WH
solutions can be studied without evolving exotic matter in certain
regions of spacetime. They concluded that WH solutions are realistic
and stable only for anisotropic matter in $f(R,T)$ gravity. Shamir
and Ahmad \cite{40} obtained the WH solutions with anisotropic
matter distribution in $f(G,T)$ gravity. They investigated some
viable regions for the presence of traversable wormhole geometries.
Sharif et al. \cite{41} analyzed static WH solutions using the
Noether symmetry technique in $f(G)$ gravity and found a stable
structure for different cases of red-shift function.

Recently, Capozziello et al. \cite{42} derived the exact traversable
WH solutions as well as stable conditions in the absence of exotic
matter in $f(R)$ theory and found that small deviation from GR give
stable solutions. De Falco et al. \cite{43} formulated the static
spherically symmetric WH solutions in the same framework. It is
interesting to mention here that for $\mathbf{T}^{2}=0$, our results
reduce to $f(R)$ gravity. We conclude that EMSG leads to the
presence of more viable and stable WH solutions for isotropic matter
configuration through Noether symmetry approach.

\end{document}